\documentclass[superscriptaddress,aps,letterpaper,10pt,twocolumn,floats,showpacs,amsmath,amsfonts,amssymb,pre,nofootinbib]{revtex4-2}
\usepackage{graphicx}
\usepackage[colorlinks=true,citecolor=blue]{hyperref}
\usepackage[caption=false]{subfig}
\usepackage{pstricks}
\usepackage{pst-node}
\usepackage{amsmath}
\usepackage{amsbsy}
\usepackage{verbatim}
\usepackage{dsfont}
\usepackage{MnSymbol}
\newcommand{\Tr}{\text{Tr}}
\DeclareMathOperator*{\argmin}{arg\,min}
\newcommand{\re}{\text{Re}}
\DeclareMathAlphabet{\mymathbb}{U}{BOONDOX-ds}{m}{n}

\begin{document}
	
\title{Dynamical signatures of discontinuous phase transitions: \\
How phase coexistence determines exponential versus power-law scaling
}
	
		\author{Krzysztof Ptaszy\'{n}ski}
 \email{krzysztof.ptaszynski@ifmpan.poznan.pl}
	\affiliation{Complex Systems and Statistical Mechanics, Department of Physics and Materials Science, University of Luxembourg, L-1511 Luxembourg, Luxembourg}
\affiliation{Institute of Molecular Physics, Polish Academy of Sciences, Mariana Smoluchowskiego 17, 60-179 Pozna\'{n}, Poland}
	
	\author{Massimiliano Esposito}
  \email{massimiliano.esposito@uni.lu}
	\affiliation{Complex Systems and Statistical Mechanics, Department of Physics and Materials Science, University of Luxembourg, L-1511 Luxembourg, Luxembourg}	
	\date{\today}
	
	\begin{abstract}
There are conflicting reports in the literature regarding the finite-size scaling of the Liouvillian gap and dynamical fluctuations at discontinuous phase transitions, with various studies reporting either exponential or power-law behavior. We clarify this issue by employing large deviation theory. We distinguish two distinct classes of discontinuous phase transitions that have different dynamical properties. The first class is associated with phase coexistence, i.e., the presence of multiple stable attractors of the system dynamics (e.g., local minima of the free energy functional) in a finite phase diagram region around the phase transition point. In that case, one observes asymptotic exponential scaling related to stochastic switching between attractors (though the onset of exponential scaling may sometimes occur for very large system sizes). In the second class, there is no phase coexistence away from the phase transition point, while at the phase transition point itself there are infinitely many attractors. In that case, one observes power-law scaling related to the diffusive nature of the system relaxation to the stationary state.
	\end{abstract}
	
	\maketitle
	
	\section{Introduction}
Dynamics of many-body systems exhibits certain characteristic features at equilibrium or nonequilibrium phase transitions, which manifest themselves already for finite system sizes. In systems described by Markovian master equations, one of the dynamical signatures of the phase transition is the closing of the \textit{Liouvillian gap} with the increase in system size. This gap is the smallest relaxation rate of the system and is given by the second dominant eigenvalue of the Liouvillian generator of the dynamics of the system~\cite{kessler2012dissipative,minganti2018spectral}. This observation has recently been generalized to the non-Markovian case~\cite{debecker2023spectral}. The Liouvillian gap closing is related to the presence of degenerate steady states of the system, which correspond to different phases that are separated by a phase transition point. When the system possesses a unique steady state for finite system sizes, the Liouvillian gap asymptotically approaches zero with increasing system size and closes exactly in the thermodynamic limit. Studies of both continuous and discontinuous phase transitions have revealed certain universal features of this asymptotic scaling behavior. In the first case, the Liouvillian gap was shown to exhibit a power-law scaling~\cite{huybrechts2020validity, wang2021dissipative}. This is related to the well-known phenomenon of critical slowing down. In particular, for classical stochastic systems, the power-law scaling of the slowest relaxation timescale has been explained by using van Kampen's expansion of the master equation~\cite{vankampen1981stochastic, gardiner2009stochastic}. The situation is less clear for discontinuous phase transitions. Such transitions often occur in the phase diagram region which admits multiple attractors of the system dynamics, e.g., local minima of the free energy functional. In this case, an exponential scaling of the Liouvillian gap has been predicted, since the slowest timescale of the system is related to stochastic switching between the attractors, which is exponentially suppressed with the system size~\cite{hanggi1984bistable, hinch2005exponentially, kurchan2009equilibrium}. This prediction was confirmed by certain numerical simulations~\cite{casteels2017critical, wang2021dissipative}. However, some other numerical studies of discontinuous phase transitions have instead observed a power-law scaling~\cite{vicentini2018critical, ferreira2019lipkin}. Furthermore, recently an experimental study of a superconducting qubit-cavity setup reported a power-law scaling of transition rates between the attractors~\cite{sett2024emergent}, which, as mentioned above, are directly related to the Liouvillian gap.

Although the Liouvillian gap is a rather abstract concept, its scaling properties manifest themselves in the corresponding scaling of measurable quantities. In particular, as discussed in Sec.~\ref{sec:spectheorfluct}, at the phase transition point, the scaling of the Liouvillian gap can determine the scaling of dynamical fluctuations of the system observables. Indeed, previous studies demonstrated the power-law scaling of fluctuations for continuous phase transitions~\cite{nguyen2018phase,remlein2024nonequilibrium} and the exponential scaling for discontinuous ones~\cite{nguyen2020exponential, fiore2021current,kewming2022diverging,remlein2024nonequilibrium}, in complete analogy to the scaling of the Liouvillian gap. (Although in quantum systems an exponential scaling of certain types of fluctuations may also be observed at continuous phase transitions due to quantum measurement effects~\cite{kewming2022diverging}.) The power-law behavior at continuous phase transitions is caused by the critical slowing down of the dynamics at the phase transition point, as recently discussed in Ref.~\cite{remlein2024nonequilibrium}. At discontinuous phase transitions, as in the case of the Liouvillian gap, the exponential scaling of fluctuations has been related to the corresponding scaling of the timescale of stochastic switching between the system attractors. Indeed, in many cases fluctuations can be quantitatively described using an effective two-state model based on this assumption~\cite{nguyen2020exponential, fiore2021current, remlein2024nonequilibrium, gopal2022large}.

In this paper, we characterize the scaling of the Liouvillian gap and of fluctuations at discontinuous phase transitions. To that end, we make use of large deviation theory~\cite{Touchette2009, FalascoReview, landauer1962fluctuations, hanggi1982stochastic, ge2009thermodynamic}. Within this framework, different equilibrium or nonequilibrium phases of the system correspond to different \textit{attractors} of the system dynamics. These attractors further correspond to the minima of the large deviation rate function describing the rare fluctuations of the system observables. The global minimum corresponds to the stationary state of the system, called the \textit{absolutely stable} attractor, while the local minima correspond to the \textit{metastable} attractors. We classify arbitrary discontinuous phase transitions into two classes. The first is associated with \textit{phase coexistence}, i.e., the presence of multiple attractors in a finite phase diagram region around the phase transition point. At the phase transition point, one of these attractors becomes absolutely stable at the expense of the other. In that case, one observes an exponential scaling of the Liouvillian gap and fluctuations, associated with stochastic switching between the attractors. In the second class of discontinuous phase transitions, there is no phase coexistence. When crossing the phase transition point, the previously existing absolutely stable attractor becomes unstable, i.e., the associated minimum of the rate function vanishes. At the phase transition point itself, there are infinitely many stable attractors, and the rate function becomes constant. In this case, the Liouvillian gap and fluctuations exhibit a power-law scaling related to the diffusive nature of the system relaxation to the stationary state. These conclusions hold unless one considers systems where, at the phase transition point, the stationary state is not unique, and thus the Liouvillian gap is closed for any system size.

We draw our conclusions from the study of several open systems. We first consider a simple model that exhibits a discontinuous phase transition at equilibrium, the molecular zipper~\cite{kittel1969phase}, which provides important analytical insights. The generality of our conclusions is then illustrated with two nonequilibrium open quantum systems: the dissipative quantum Ising model~\cite{wang2021dissipative} and the anisotropic LMG model~\cite{lee2014dissipative,debecker2023spectral}. We also briefly discuss the squeezed decay model~\cite{agarwal1989nonequilibrium, agarwal1990cooperative,barberena2023critical}, where the Liouvillian gap is closed for any finite system size at the phase transition point.

The paper is organized as follows. In Sec.~\ref{sec:spectheordyn} we discuss the spectral decomposition of the Liouvillian and define the concept of Liouvillian gap. In Secs.~\ref{sec:spectheorfluct} and~\ref{sec:disc} we briefly review the spectral theory of fluctuations and the large deviation theory of discontinuous phase transitions. Secs.~\ref{sec:zipp}--\ref{sec:sdm} present results for the molecular zipper, open quantum Ising model, anisotropic LMG model, and the squeezed decay model, respectively. Finally, Sec.~\ref{sec:concl} draws the conclusions from our results. Appendix~\ref{app:derexpgap} presents the derivation of the Liouvillian gap formula for the molecular zipper in the phase coexistence region.
	
	\section{Spectral properties and gap of the Liouvillian} \label{sec:spectheordyn}
Let us first briefly review the description of the open system dynamics in terms of spectral properties of the Liouvillian and then introduce the concept of Liouvillian gap. To simplify the discussion, we start from the classical case. Consider a Markovian system consisting of $M$ discrete states $i \in \{1,\ldots,M \}$ with probabilities $p_i$. The dynamics of state probabilities can be described by the master equation 
\begin{align} \label{clmast}
	\dot{\mathbf{p}}=W \mathbf{p},
\end{align}
where $\mathbf{p}=(p_1,\ldots,p_M)^T$ is the vector of state probabilities, and $W$ is the rate matrix, also called the Liouvillian. The off-diagonal elements of the rate matrix $W_{ij}$ $(i \neq j)$ are the transition rates from the $j$th to the $i$th state, while the diagonal elements $W_{ii}=-\sum_{j \neq i} W_{ji}$ are responsible for the probability conservation. In our paper, we further focus on the systems whose rate matrix is irreducible, i.e., every state $i$ can be reached from every other state $j$ by a sequence of Markov jumps. In such a case, by virtue of the Perron-Frobenius theorem, for finite state spaces, the system has a unique steady state $\mathbf{p}_{ss}$ (with all state probabilities being strictly positive), corresponding to the solution of the equation
\begin{align} \label{classtst}
	W \mathbf{p}_{ss}=0.
\end{align}
This description can be generalized to open quantum systems with dynamics described by the Lindblad equation (we take $\hbar=1)$
\begin{align} \label{lindblad}
	\dot{\rho}=-i[H,\rho]+\sum_k \gamma_k \mathcal{D}[L_k](\rho),
\end{align}	
where the two successive terms on the right-hand side correspond to the unitary and dissipative dynamics of the system, respectively, $H$ is the effective Hamiltonian of the system, $\gamma_k$ are the jump rates, and
\begin{align}
	\mathcal{D}[L_k](\rho) \equiv L_k \rho L_k^\dagger-\frac{1}{2} \left\{L_k^\dagger L_k,\rho \right\},
\end{align}
where $L_k$ are the Lindblad jump operators. The Lindblad equation is often written in short as
\begin{align}
	\dot{\rho}=\mathcal{L} \rho,
\end{align}
where generators of unitary and dissipative dynamics are merged into one superoperator, called the Liouvillian.

We now rewrite the Lindblad equation~\eqref{lindblad} in the matrix form analogous to the classical master equation~\eqref{clmast}. To this end, we employ the Liouville space representation~\cite{fano1963pressure,mukamel1999principles}. Within this approach, the $d \times d$ density matrix $\rho$ (where $d$ is the Hilbert space dimension of the system) is represented by the $d^2$ column vector $|\rho \rrangle$,
\begin{align}
	\rho \rightarrow |\rho \rrangle,
\end{align}
such that element $\rho_{ij}$ of the density matrix corresponds to $[i+(j-1)d]$th element of the vector $|\rho \rrangle$. Operators in the Liouville space can be represented as
\begin{align}
	\hat{A} \rightarrow \bar{A} = \mathds{1} \otimes \hat{A},
\end{align}
and their expected values can be calculated as	
\begin{align}
	\Tr(	\hat{A} \rho) = \llangle \mathds{1}|\bar{A}|\rho \rrangle,
\end{align}
where $\llangle \mathds{1}|$ is the row vector with elements equal to 1 at positions $i+(i-1)d$, and 0 otherwise. The dynamics of the vector $|\rho \rrangle$ is given by the equation
\begin{align}
	\frac{d}{dt} |\rho \rrangle=\bar{\mathcal{L}} |\rho \rrangle,
\end{align}	
where 
\begin{align} \nonumber
	&\bar{\mathcal{L}} =-i \left( \mathds{1} \otimes H-H^T \otimes \mathds{1} \right) \\ &+\sum_{k} \gamma_k \left[L_k^* \otimes L_k - \frac{1}{2}\mathds{1} \otimes L_k^\dagger L_k - \frac{1}{2} \left(L_k^\dagger L_k \right)^T \otimes \mathds{1} \right]
\end{align}
is the matrix representation of the Liouvillian~\cite{machnes2014surprising, amshallem2015approaches, uzdin2016speed,minganti2018spectral}. Note that $^*$ here denotes the complex conjugate, while $^\dagger$ is the Hermitian conjugate. Analogously to the classical case, the steady state is given by the solution of the equation
\begin{align} \label{eq:quantss}
	\bar{\mathcal{L}}|\rho_{ss} \rrangle=0.
\end{align}
For finite Hilbert spaces, the steady state of the system is unique and faithful (spanned over the whole Hilbert space of the system) if and only if its dynamics is Davies-irreducible, i.e., when the condition
\begin{align} \label{eq:daviesirr}
  \forall_{\hat{\rho}, t}:  e^{\mathcal{L} t} \hat{P} \hat{\rho} \hat{P}=\hat{P} \left(e^{\mathcal{L} t} \hat{P} \hat{\rho} \hat{P} \right) \hat{P},
\end{align}
where $\hat{P}$ is a Hermitian projection operator fulfilling $\hat{P} \hat{P}=\hat{P}$, is satisfied only for $\hat{P}=\mathds{1}$ and $\hat{P}=\mymathbb{0}$~\cite{davies1970quantum, zhang2024criteria}. In other words, there is no subspace of the Hilbert space such that, for an initial state confined to that subspace, the final state will also be confined to that subspace. An example where this is not true is considered in Sec.~\ref{sec:sdm}.

We now discuss how dynamics of the system can be described in terms of the spectral decomposition of the Liouvillian. For a time-independent Liouvillian, the state of the system at time $t$ can be written as
\begin{align} \label{eq:dynvecgen}
	|\rho(t) \rrangle = e^{\bar{\mathcal{L}} t} |\rho(0) \rrangle,
\end{align}
where $|\rho(0) \rrangle$ is the initial state and $e^{\bar{\mathcal{L}} t}$ is the propagator of the dynamics. It can be spectrally decomposed as\footnote{Here we assume that the Liouvillian is diagonalizable, and thus its right and left eigenvectors form a complete basis. The opposite situation, when the Liouvillian becomes defective and some eigenvectors become degenerate, may occur for certain points in the parameter space, called the exceptional points~\cite{minganti2019quantum}. However, such points are rare and are not relevant in the context of the present paper.}~\cite{vankampen1981stochastic,hanggi1982stochastic,minganti2018spectral}
\begin{align} \label{eq:propag}
	e^{\bar{\mathcal{L}}t}=	|\rho_{ss} \rrangle\llangle \mathds{1}|+\sum_{j \neq 0} e^{\lambda_j t}|x_j \rrangle \llangle y_j |,
\end{align}
where $|x_j \rrangle$ and $\llangle y_j |$ are right and left eigenvectors of the Liouvillian corresponding to the eigenvalue $\lambda_j$:
\begin{align}
	\bar{\mathcal{L}}|x_j \rrangle =\lambda_j |x_j \rrangle, \quad \llangle y_j |\bar{\mathcal{L}} =\lambda_j \llangle y_j |.
\end{align}
The eigenvectors are normalized as $\llangle y_j|x_j \rrangle=1$. The eigenvalues of the Liouvillian are here ordered in the decreasing order of their real parts: $\re(\lambda_0) \geq \re(\lambda_1) \geq \ldots \geq \re(\lambda_{d^2-1})$. When the steady state is unique, the Liouvillian has a single dominant eigenvalue $\lambda_0$ equal to 0, and the associated right and left eigenvectors correspond to $|\rho_{ss} \rrangle$ and $\llangle \mathds{1}|$, respectively. Consequently, Eq.~\eqref{eq:dynvecgen} can be rewritten as
\begin{align} \label{eq:dynvecdec}
	|\rho(t) \rrangle = |\rho_{ss} \rrangle +\sum_{j \neq 0} e^{\lambda_j t}|x_j \rrangle \llangle y_j |\rho(0) \rrangle.
\end{align}
Now, we can see that the real parts of the eigenvalues $\lambda_i$ ($i>0$) describe the relaxation rates of the system to the stationary state. In particular, the absolute value of the slowest relaxation rate is known as the \textit{Liouvillian gap}
\begin{align} \label{liovgap}
	\lambda \equiv |\re(\lambda_1)|.
\end{align}
When the steady state is unique, the Liouvillian gap is strictly positive. Instead, when the steady state is $n$-fold degenerate, the Liouvillian has $n$ dominant eigenvalues equal to 0~\cite{hanggi1982stochastic,minganti2018spectral}, and thus the Liouvillian gap is closed ($\lambda=0$). This may occur in two situations:
\begin{itemize}
\item When the system dynamics is not irreducible for finite system sizes (see discussion below Eqs.~\eqref{clmast} and~\eqref{eq:quantss}).
\item When the system dynamics is irreducible for finite system sizes, but the Liouvillian gap closing and the steady state degeneracy emerges in the thermodynamic limit of an infinite system size~\cite{minganti2018spectral} (where the Perron-Frobenius theorem and its quantum generalizations~\cite{davies1970quantum,zhang2024criteria} do not apply, as they are formulated for finite-dimensional Liouvillians).
\end{itemize}
In our work, we mostly focus on the latter case, though the former case is briefly reviewed in Sec.~\ref{sec:sdm}.

Finally, we note that despite corresponding to the slowest relaxation rate of non-stationary eigenvectors of the Liouvillian, the Liouvillian gap cannot always be simply identified with the slowest relaxation timescale of the system, as the latter may result from the interplay of several relaxation processes~\cite{mori2020resolving,haga2021liouvillian,bensa2021fastest,lee2023anomalously,mori2023symmetrized}.

	\section{Spectral theory of fluctuations} \label{sec:spectheorfluct}
As spectral decomposition of the Liouvillian fully characterizes the system dynamics, it also provides a convenient way to determine the properties of dynamical fluctuations of system observables described by Hermitian operators $\hat{o}$~\cite{scarlatella2019spectral, hanggi1982stochastic, vankampen1981stochastic}. They can be characterized by means of the noise spectral density~\cite{clerk2010introduction}
\begin{align} \label{noisespec}
	S_{oo}(\omega) \equiv \int_{-\infty}^\infty dt e^{i \omega t} \langle \hat{o}(t) \hat{o}(0) \rangle,
\end{align}
where $\langle \hat{o}(t) \hat{o}(0) \rangle$ is the two-time correlation function. For Markovian open quantum systems, this function can be calculated using the quantum regression theorem~\cite{lax1963formal}
\begin{align} \label{quantreg}
	\langle \hat{o}(t) \hat{o}(0) \rangle=\Tr \left(\hat{o} e^{\mathcal{L} t} \hat{o} \rho_{ss} \right),
\end{align}
which is applicable for $t \geq 0$. The correlation function for negative times is given by the relation $\langle \hat{o}(-t) \hat{o}(0) \rangle=\langle \hat{o}(t) \hat{o}(0) \rangle^*$~\cite{hanggi1982stochastic}. The noise spectral density can thus be calculated as
\begin{align} \label{noisespecmark}
	S_{oo}(\omega)=\int_{0^+}^\infty dt e^{i \omega t} \Tr \left(\hat{o} e^{\mathcal{L} t} \hat{o} \rho_{ss} \right)+\text{c.c.}
\end{align}
Using Eq.~\eqref{eq:propag}, the two-time correlation function can be written as
\begin{align}
	\Tr \left(\hat{o} e^{\mathcal{L} t} \hat{o} \rho_{ss} \right)=\llangle \mathds{1}| \bar{o}|\rho_{ss} \rrangle^2+\sum_{j \neq 0} e^{\lambda_j t} \llangle \mathds{1}|\bar{o}|x_j \rrangle \llangle y_j |\bar{o}|\rho_{ss} \rrangle.
\end{align}
Inserting this expression into Eq.~\eqref{noisespecmark}, we obtain the spectral decomposition of the noise spectral density~\cite{hanggi1982stochastic,vankampen1981stochastic}
\begin{align} \label{eq:specdecompnoise}
	S_{oo}(\omega)=-\sum_{j \neq 0} \frac{1}{i\omega+\lambda_j} \mathcal{A}^{(j)}_{oo}+\text{c.c.},
\end{align}
where we define the overlap functions
\begin{align} \label{eq:overlap}
\mathcal{A}^{(j)}_{oo} \equiv \llangle \mathds{1}| \bar{o}|x_j \rrangle \llangle y_j |\bar{o}|\rho_{ss} \rrangle.
\end{align}
In particular, zero-frequency noise can be expressed as
\begin{align} \label{zerofreqnoise}
	S_{oo}(0)=-\sum_{j \neq 0} \frac{1}{\lambda_j}\mathcal{A}^{(j)}_{oo}+\text{c.c.}
\end{align}

This expression enables us to relate the finite-size scaling of fluctuations in systems exhibiting phase transitions to the scaling of the Liouvillian gap. Let us focus on the case where the eigenvalue $\lambda_1$ is real (which is the case for the models considered in this paper, and other discontinuous phase transition models known to us). It can now be argued that when the Liouvillian gap closes ($\lambda \approx 0$), which occurs at the phase transition point, fluctuations are dominated by the term related to $\lambda$:
\begin{align}
	S_{oo}(0) \approx \frac{1}{\lambda} 2 \text{Re} \: \mathcal{A}_{oo}^{(1)}.
\end{align}
Thus, either exponential or power-law closing of the Liouvillian gap should be reflected in the exponential or power-law divergence of the noise spectral density. More precisely, this is true provided that the overlap function $\mathcal{A}_{oo}^{(1)}$ is non-vanishing; an example where this is not the case is presented in Sec.~\ref{subsec:isingscalgc}.

Eq.~\eqref{zerofreqnoise} can also be rewritten in a concise form~\cite{landi2023current}
\begin{align}
S_{oo}(0) = -2 \re \llangle  \mathds{1}| \bar{o} \bar{\mathcal{L}}^D \bar{o}|\rho_{ss} \rrangle,
\end{align}
where $\bar{\mathcal{L}}^D=\sum_{j=1}^{d^2-1} \lambda_j^{-1} |x_j \rrangle \llangle y_j|$ is the Drazin inverse of the Liouvillian~\cite{crook2018drazin}. This formula is convenient for numerical calculations. Finally, while we focus on system observables, an analogous spectral decomposition can be applied for current fluctuations~\cite{esposito2007fluctuation,flindt2008counting,walldorf2020noise}; see Ref.~\cite{landi2023current} for a recent tutorial paper.

\section{Discontinuous phase transitions} \label{sec:disc}
 %
\begin{figure}
	\centering
	\includegraphics[width=0.95\linewidth]{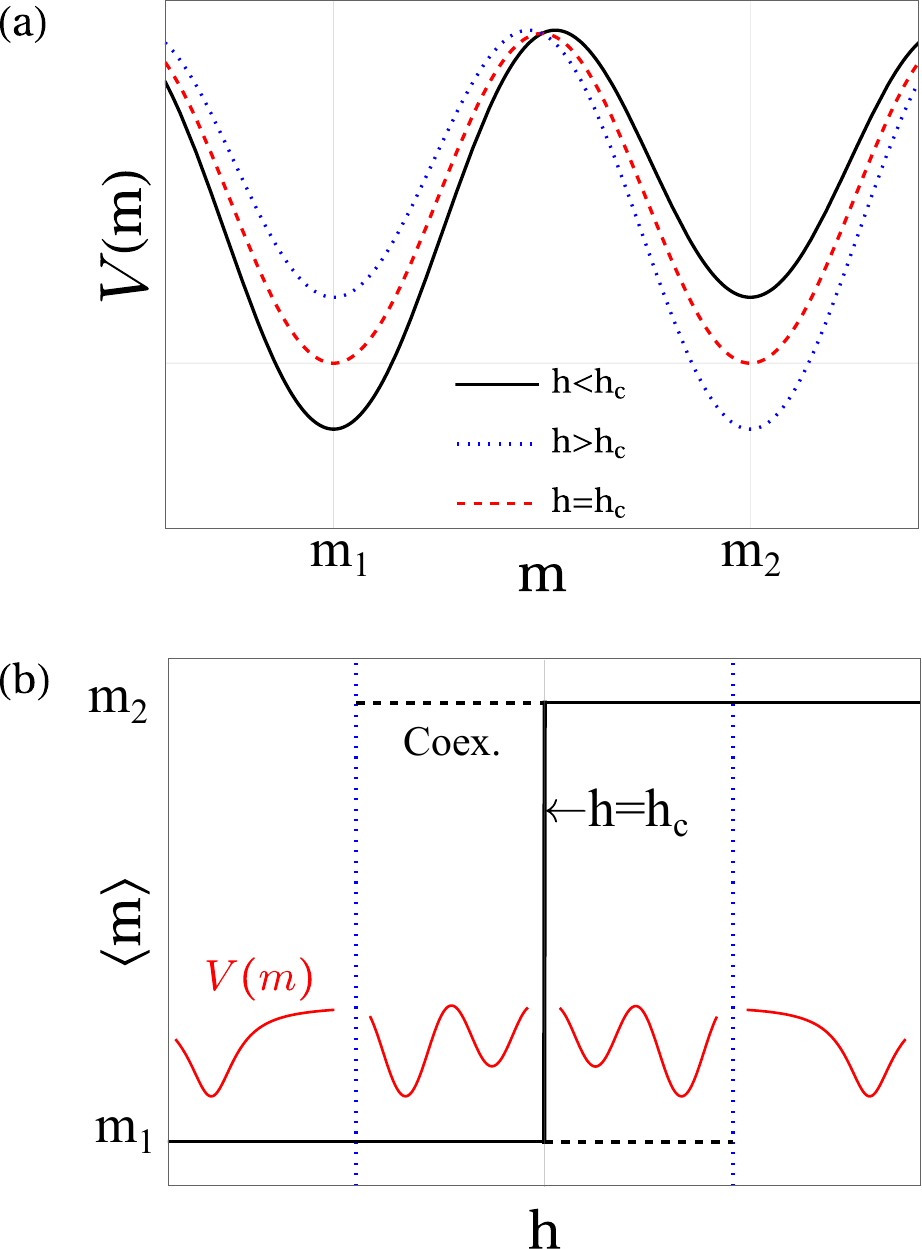}
	\caption{(a) Schematic presentation of the steady-state rate function in the phase coexistence region for different values of $h$. (b) Schematic presentation of the discontinuous phase transition for the observable $m$ (black solid line). The black dashed lines represent the metastable attractors, and the vertical blue dotted lines denote the borders of the phase coexistence region (Coex.). Small inset plots in the bottom of the graph (red solid lines) present the behavior of the rate function $V(m)$ in different phase diagram regions.}
	\label{fig:ratefunschem}
\end{figure}
%
Before focusing on specific models, let us first briefly review a general theory of discontinuous phase transitions, applicable both in and out of equilibrium. Consider a generic system characterized with a probability distribution $P(m)$ of some intensive observable $m$ (e.g., magnetization or concentration of a chemical species). Let us also take the size of the system (e.g., number of spins or volume) to be parameterized by a scaling parameter $N$. In many physical situations, for large system sizes, the probability distribution $P(m)$ takes an asymptotic large deviation form~\cite{Touchette2009,FalascoReview,landauer1962fluctuations,hanggi1982stochastic,ge2009thermodynamic}
\begin{align} \label{eq:largedev}
P(m) \asymp e^{-N V(m)},
\end{align}
where $V(m)$ is the scale-independent rate function. In particular, for equilibrium systems, the rate function may correspond to the free energy density functional. The formula above implies that for a large system size the probability distribution $P(m)$ becomes narrowly peaked around the position of the global minimum of the rate function. Then, the average value of the observable $m$ corresponds to the position of this minimum:
\begin{align}
\langle m \rangle \asymp \argmin_{m} V(m).
\end{align}
(For simplicity, here we take the global minimum to be unique; one needs to be more careful in the case of degeneracy). In some physical situations, the rate function can have several minima, which
is schematically represented in Fig.~\ref{fig:ratefunschem}~(a). This is called \textit{multistability} or \textit{phase coexistence}. The minima of the rate function then correspond to \textit{attractors} of the system dynamics~\cite{hanggi1982stochastic,ge2009thermodynamic,FalascoReview}. In particular, the global minimum corresponds to the \textit{absolutely stable} attractor, which tends to be occupied with probability 1 in the thermodynamic limit $N \rightarrow \infty$. The local minima correspond instead to the \textit{metastable} attractors of the system. For finite system sizes, the metastable attractors tend to relax over time to the absolutely stable attractor. However, the relaxation timescales (called the lifetimes of metastable states) increase with the system size $N$ and diverge for $N \rightarrow \infty$~\cite{hanggi1982stochastic,hanggi1984bistable,hinch2005exponentially,kurchan2009equilibrium}.

Let us further assume that the rate function $V(m)$ depends on some control parameter $h$ (e.g., magnetic field, temperature, pressure, or concentration of other chemical species). The range of parameter $h$ in which the system possesses multiple attractors is called the \textit{phase coexistence region}. The parameter $h$ further determines which of the minima of the rate function is global, and thus which attractor is absolutely stable. Specifically, in the case presented in Fig.~\ref{fig:ratefunschem}~(a), the average value of the observable $m$ takes the value $m_1$ ($m_2)$ for $h<h_c$ ($h>h_c)$, where $h_c$ is called the critical value of the parameter $h$, or the phase transition point. At $h=h_c$, both minima have the same value and thus the absolutely stable attractor becomes degenerate. Consequently, when the parameter $h$ is tuned, in the thermodynamic limit $N \rightarrow \infty$ the observable $m$ exhibits a discontinuous phase transition at $h=h_c$, that is, its average value exhibits a discontinuous jump from $m_1$ to $m_2$. This is schematically illustrated in Fig.~\ref{fig:ratefunschem}~(b). We note that here we refer to the behavior of the stationary value of the observable $m$, assuming that the system relaxes to the stationary state. When the parameter $h$ is swept fast compared to the relaxation time of the metastable attractor, the system exhibits rather a hysteretic behavior, staying in the metastable attractor after crossing the phase transition point.

However, while the picture presented here is quite common in physical systems, in some instances a discontinuous phase transition point is not surrounded by the phase coexistence region. Instead, the phase transition point separates the phase diagram regions with distinct attractors (i.e., when crossing the phase transition point from one side, all previously existing attractors vanish)~\cite{kittel1969phase,ferreira2019lipkin,agarwal1989nonequilibrium,agarwal1990cooperative,barberena2023critical,huber2020nonequilibrium,janssen2016first}. In the examples considered in this paper, the rate function $V(m)$ becomes constant, and thus there are infinitely many attractors (i.e., every value of $m$ corresponds to some attractor). As we will see, both scenarios (with and without phase coexistence) are associated with a qualitatively different behavior of the Liouvillian gap and fluctuations. In the former case, their finite-size scaling is dominated by stochastic switching between attractors, which is exponentially suppressed with the system size. Instead, in the latter case, this scaling is dominated by the diffusion between attractors and thus exhibits a power-law behavior.
	
	\section{Molecular Zipper model} \label{sec:zipp}
	\subsection{Equilibrium thermodynamics of the model}	
 %
\begin{figure}
	\centering
	\includegraphics[width=0.9\linewidth]{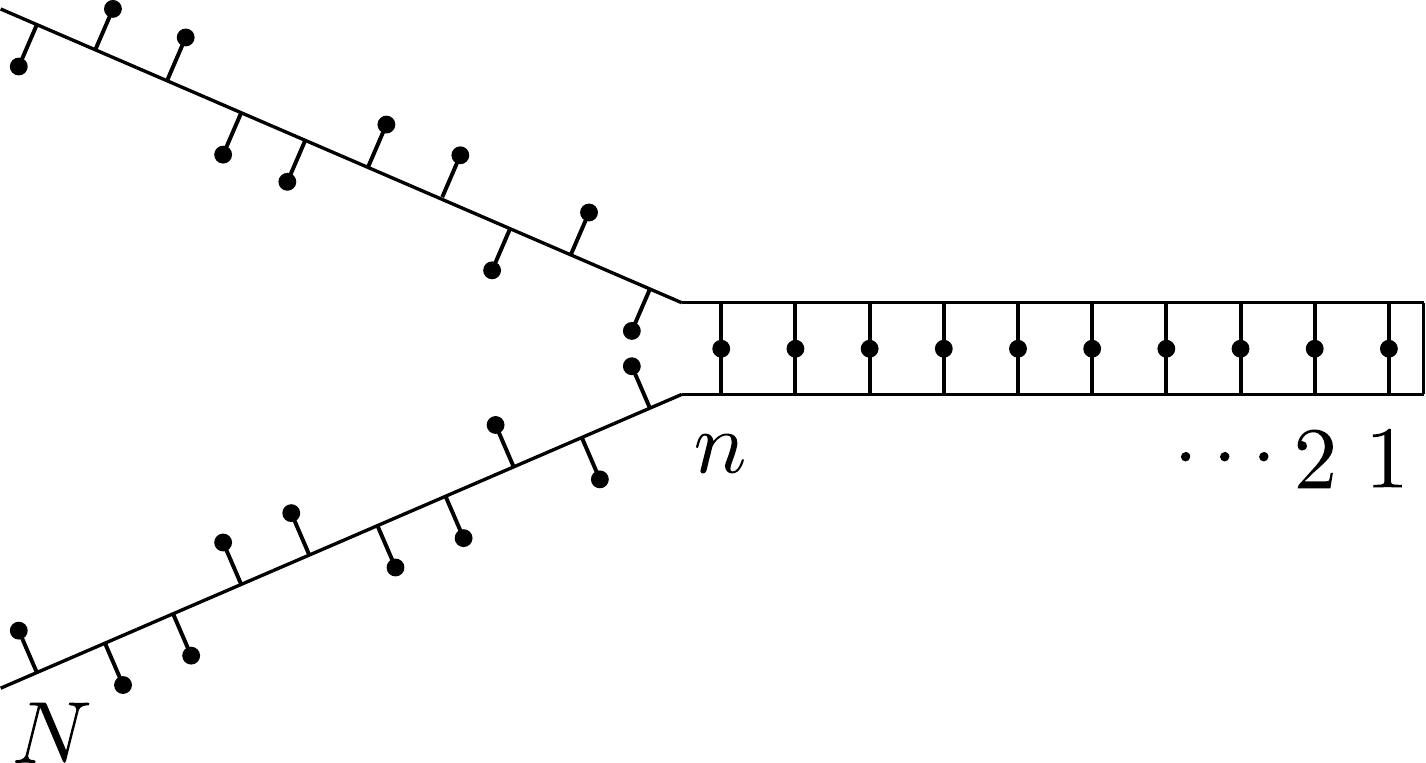}
	\caption{Scheme of the molecular zipper with $n=10$ closed links and $N-n=10$ open links. The scheme corresponds to the case of $g=2$, with open links taking two orientations: ``inward'' and ``outward''.}
	\label{fig:zipperscheme}
\end{figure}
%
Let us now analyze the finite-size scaling of the Liouvillian gap and fluctuations in an equilibrium system exhibiting a discontinuous phase transition. We consider a generalization of the molecular zipper proposed by Kittel~\cite{kittel1969phase}, a toy model of the unwinding transition in DNA molecules. Due to its simplicity and analytic tractability, it has received interest in other physical contexts, such as large deviation theory~\cite{touchette2010, deger2018lee}, nonequilibrium dynamics and thermodynamics~\cite{holubec2012dynamics}, or melting of thin films~\cite{abdullah2015zipper}. We emphasize, however, that our intention here is not to describe the behavior of any specific real-world system. Instead, our goal is to develop an analytically tractable toy model of discontinuous phase transitions that will shed light on the phenomena observed in the microscopically justified but much more complex models discussed in the following sections.

The model consists of a double-stranded macromolecule, rigidly connected at one end, stabilized by $N$ parallel links that can be either closed or open (Fig.~\ref{fig:zipperscheme}). The $i$th link can close only if the $i-1$ preceding links are also closed. Closing of the $i$th link decreases the energy of the system by $\epsilon_i$. (This generalizes the original model, where all energies $\epsilon_i$ were equal to each other\footnote{The original model has also used a different convention, where an open link was characterized with a positive energy $\epsilon$. Our convention is more convenient in the case where the energies of closed links can vary.}.) The energy of the system with $n$ closed links is then equal to
\begin{align}
E_n=-\sum_{i=1}^n \epsilon_i.
\end{align}

It is also assumed that the link can be opened in $g$ different energy-degenerate ways (e.g., open links can be oriented in different directions). Thus, the system with $n$ closed links corresponds to $g^{(N-n)}$ different microscopic configurations of the system. Consequently, the Boltzmann entropy of the system with $n$ closed links is equal to 
\begin{align}
S_n=k_B \ln g^{(N-n)}=(N-n) k_B \ln g.
\end{align}

The system is also coupled to an ideal thermal bath with a temperature $T$. For convenience of notation, we often also use the inverse temperature $\beta=1/(k_B T)$. We define the free energy functional of the system $F_n \equiv E_n-T S_n$:
\begin{align} \label{eq:freeenfunzippdisc}
	F_n=-\sum_{i=1}^n \epsilon_i -(N-n) k_B T \ln g.
\end{align}
The probability that $n$ links are closed is then given by Boltzmann distribution
\begin{align} \label{eq:probn}
p_n = Z^{-1} e^{-\beta F_n}
\end{align}
where $Z=\sum_{m=0}^N e^{-\beta F_m}$ is the partition function.

We now parameterize the link energies $\epsilon_i$ such that the model exhibits the large deviation behavior of probabilities $p_n$, given by Eq.~\eqref{eq:largedev}. To do that, we take the energies $\epsilon_i$ to be given by a scale-invariant function $f(x)$:
\begin{align}
	\epsilon_i=f\left (\frac{i}{N} \right).
\end{align}
We then consider the limit of the large system size $N$ and parameterize $n$ with the rescaled variable $q=n/N \in[0,1]$. We define the free energy density functional
\begin{align} \label{eq:freeenfun}
\mathcal{F}(q) \equiv \lim_{N \rightarrow \infty} \frac{F_n}{N} \quad \text{for} \quad q=\frac{n}{N}.
\end{align}
Taking a continuous limit of Eq.~\eqref{eq:freeenfunzippdisc}, the free energy density functional can be expressed as
\begin{align} \label{eq:freeendenzippgen}
\mathcal{F}(q) =-\int_0^q f(x) dx-(1-q) k_B T \ln g.
\end{align}

Let us now consider the equilibrium behavior of the system. In the large $N$ limit, the absolutely stable attractor, corresponding to the equilibrium state, can be found by minimizing the free energy density functional $\mathcal{F}(q)$. Indeed, using Eqs.~\eqref{eq:probn} and~\eqref{eq:freeenfun}, the probabilities that $n$ links are closed scale in the large $N$ limit as
\begin{align} \label{eq:largedevzipp}
p_n \propto e^{-N \beta \mathcal{F}(n/N)} \equiv e^{-N V(n/N)},
\end{align}
where $V(q)=\beta \mathcal{F}(q)$ is the rate function defined in Sec.~\ref{sec:disc}. Consequently, using theory from Sec.~\ref{sec:disc}, the equilibrium average number of the closed links takes an asymptotic form
\begin{align}
\langle n \rangle_\text{eq} \asymp N q_\text{eq},
\end{align}
where $q_\text{eq}$ is the value of $q$ minimizing $\mathcal{F}(q)$:
\begin{align} \label{eq:qeq}
q_\text{eq} = \operatorname*{arg\,min}_{q \in [0,1]} \mathcal{F}(q).
\end{align}
We note that minimization is restricted to the domain of $q$, that is, the interval $[0,1]$.

We then focus on a particular model where the energies of closed links are given by a linear function:
\begin{align} \label{eq:fx}
	f(x)=\epsilon(1+2ax),
\end{align}
with $a \geq 0$. Using Eq.~\eqref{eq:freeendenzippgen}, the free energy density functional becomes a quadratic function of $q$:
\begin{align} \label{eq:freeendenzippspec}
	\mathcal{F}(q) =(k_B T \ln g-\epsilon)q-\epsilon a q^2-k_B T \ln g.
\end{align}
Minimizing $\mathcal{F}(q)$, we find that in the thermodynamic limit the system exhibits a discontinuous phase transition: the zipper is closed ($q_\text{eq}=1$) below the critical temperature
\begin{align} \label{eq:crittempzipp}
	T_c&=\frac{(1+a)\epsilon}{k_B \ln g},
\end{align}
while it is open ($q_\text{eq}=0$) above this temperature. Furthermore, in a finite temperature window below and above $T_c$ one observes the phase coexistence: the free energy density functional has two minima at $q=0$ and $q=1$. As discussed in Sec.~\ref{sec:disc}, the global minimum corresponds to the absolutely stable attractor (i.e., the equilibrium state), while the local minimum corresponds to the metastable attractor. The lower and upper borders of the temperature window of the phase coexistence read as
\begin{align}
        T_1=\frac{\epsilon}{k_B \ln g}, \quad
	T_2=\frac{\epsilon(1+2a)}{k_B \ln g}.
\end{align}
\begin{figure}
	\centering
	\includegraphics[width=0.9\linewidth]{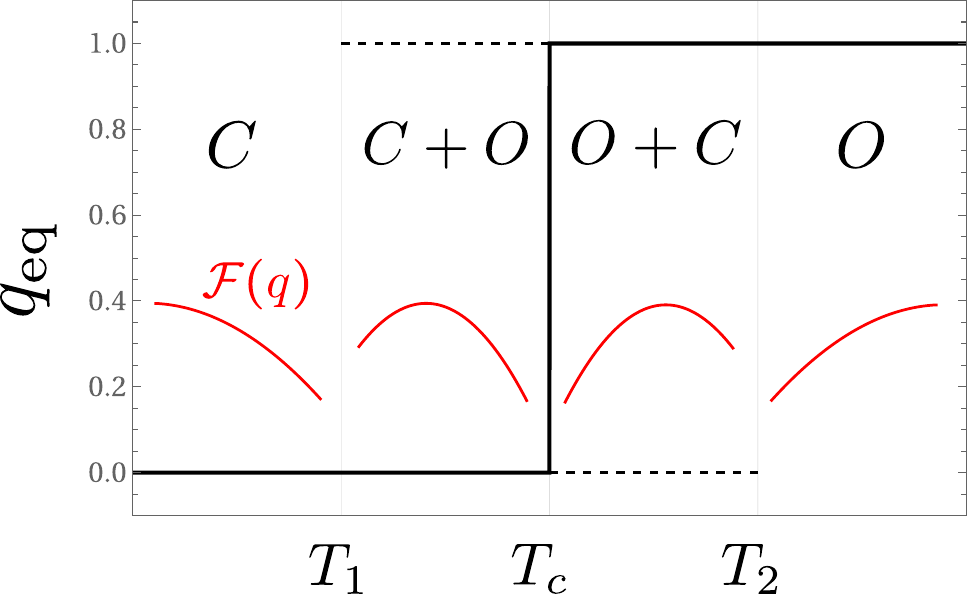}
	\caption{Schematic temperature-dependence of the order parameter $q_\text{eq}$ (black solid line). The dashed lines represent the metastable attractors related to local minima of the free energy density functional. The letter $C$ ($O$) denotes the temperature region in which only the closed (open) attractors exists. The letters $C+O$ ($O+C$) denote the region in which the closed (open) attractor is absolutely stable, while the open (closed) attractor is metastable. Small inset plots in the bottom of the graph (red solid lines) present the behavior of the free energy density functional $\mathcal{F}(q)$ (for $q \in [0,1]$) in different phase diagram regions.}
	\label{fig:phasediagzipp}
\end{figure}
The schematic presentation of the thermodynamic behavior of the order parameter $q_\text{eq}$ is presented in Fig.~\ref{fig:phasediagzipp}. As one may observe, for $T_1<T<T_c$ ($T_c<T<T_2$) the thermodynamic equilibrium corresponds to the closed (open) state, while the open (closed) state is metastable. In contrast, for $T<T_1$ ($T>T_2$) the free energy density functional has a single minimum that corresponds to the closed (open) attractor.

We recall that the original molecular zipper model corresponds to the case of $a=0$. In this case, $T_c=T_1=T_2$, and thus the phase coexistence region vanishes. Furthermore, at the phase transition point, the probability distribution $\{ p_n \}_n$ becomes uniform, and thus the free energy density functional $\mathcal{F}(q)$ and the rate function $V(q)=\beta \mathcal{F}(q)$ become constant. As we later show, this has huge consequences for the scaling of the Liouvillian gap and fluctuations.

\subsection{Master equation}
Let us now present the dynamical description of the model using the Markovian master equation. In fact, there is an infinite number of dynamical models which are consistent with the equilibrium description presented in the previous subsection. For the sake of simplicity, we use the following assumption: As mentioned, every link can be opened in $g$ degenerate ways. We assume that when the $i$th link opens, each of these degenerate states is generated with equal probability. Analogously, for every state of an open link, the transition rate to close the link is the same. Then, the system can be described using a coarse-grained mesoscopic master equation~\cite{herpich2020njp} for probabilities $p_i$ that $i$ links are closed,
\begin{align}
\dot{p}_i=\sum_{j} \left( W_{ij} p_j -W_{ji} p_i\right),
\end{align}
where $W_{ij}$ is the transition from the state with $j$ closed links to the state with $i$ closed links. We recall that each state with $i$ closed links, associated with probability $p_i$, corresponds to $g^{(N-i)}$ equally probable microscopic configurations of the system. To provide consistency with thermodynamics, the transition rates must satisfy the detailed balance condition~\cite{herpich2020njp}\footnote{This condition can be generalized to a multiple-bath nonequilibrium scenario, where a similar \textit{local} detailed balance condition holds~\cite{herpich2020njp}}.
\begin{align}
	\ln \frac{W_{i-1,i}}{W_{i,i-i}}=\beta (F_i-F_{i-1}),
\end{align}
where $F_i$ is the free energy functional defined in Eq.~\eqref{eq:freeenfunzippdisc}. Therefore, the ratio of rates related to the opening and closing of the $i$th link is related to the free energy cost to close the $i$th link. As a specific choice of transition rates that fulfill this condition, we use Arrhenius rates
\begin{align}
	\begin{cases}
		W_{i,i-1} &=\Gamma \exp(\beta \epsilon_{i}/2) \quad (i=1,\ldots,N), \\
		W_{i-1,i} &=g \Gamma \exp(-\beta \epsilon_{i}/2) \quad (i=1,\ldots,N), \\
		W_{ij} &=0 \quad \text{otherwise for $i \neq j$}.
	\end{cases}
\end{align}
Recall that the diagonal rate matrix elements read $W_{ii}=-\sum_{j \neq i} W_{ji}$.

\subsection{Liouvillian gap scaling}
\subsubsection{$a=0$: power-law scaling}
Let us now consider the finite-size scaling of the Liouvillian gap at the critical temperature $T=T_c$. We first focus on the case of $a=0$, where all energies $\epsilon_i$ are equal to each other. As mentioned, in this case the phase coexistence region vanishes. At the critical temperature $T_c$ all non-zero transition rates are equal and can be expressed as
\begin{align}
W_{i,i+1}=W_{i+i,i}=\Gamma \sqrt{g}.
\end{align}
Spectral decomposition of the rate matrix with such transition rates is known in the theory of Laplacian matrices. Specifically, the eigenvalues of the rate matrix read~\cite{britanak2010discrete}
\begin{align} \label{eq:eigzipper}
	\lambda_j=-4\Gamma \sqrt{g} \sin^2 \left[\frac{\pi j}{2(N+1)} \right].
\end{align}

\begin{figure}
	\centering
	\includegraphics[width=0.9\linewidth]{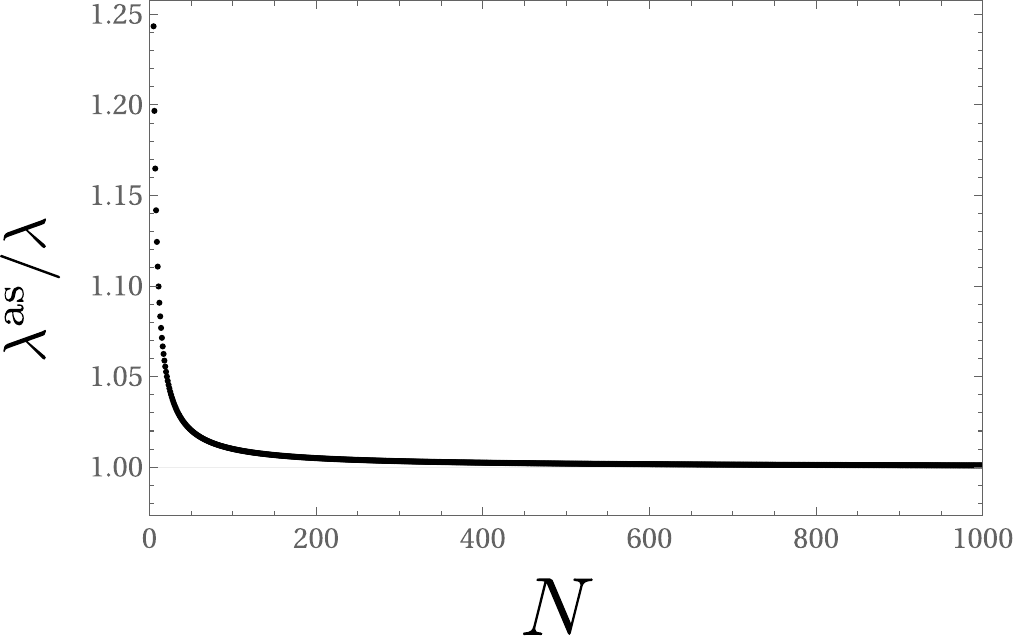}
	\caption{The ratio of the asymptotic Liouvillian gap $\lambda^\text{as}$, given by Eq.~\eqref{eq:gapapproxa0}, and the exactly calculated Liouvillian gap $\lambda$ for the molecular zipper with $a=0$.}
	\label{fig:gapapproxa0}
\end{figure}
%
Using the Taylor expansion and focusing on the leading order of $N$, it is now clear that in the large $N$ limit the Liouvillian gap exhibits an asymptotic power-law scaling:
\begin{align} \label{eq:gapapproxa0}
	\lambda \asymp \lambda^\text{as}=\frac{\pi^2 \Gamma \sqrt{g}}{N^2}.
\end{align}
Fig.~\ref{fig:gapapproxa0} presents the ratio of this asymptotic power-law formula and the exact result. We can observe that the Liouvillian gap follows the asymptotic $\propto N^2$ behavior very well for large enough system sizes $N \gtrapprox 400$.

\textit{Diffusive mechanism of power-law scaling.} Let us now rationalize this result using another approach. In the large $N$ limit, instead of the discrete probability distribution $\{ p_n \}_n$, one can use the probability density $\rho(q)=N p_n$ with $q=n/N$. Using Kramers-Moyal expansion of the master equation~\cite{kramers1940brownian,moyal1949stochastic}, the equation of motion for $\rho(q)$ can be written as~\cite{FalascoReview}
\begin{align}
\partial_t \rho(q)=-\frac{\partial}{\partial q} \left[\mu(q) \rho(q) \right]+\frac{\partial^2}{\partial q^2} \left[D(q) \rho(q) \right]+O(N^{-3}),
\end{align}
where $\mu(q)=[w_+(q)-w_-(q)]/N$ is the drift term and $D(q)=[w_+(q)+w_-(q)]/(2N^2)$ is the diffusion coefficient. The functions $w_\pm (q)$ are here the transition rates $W_{i\pm1,i}$ for $i=qN$ and $N \rightarrow \infty$:
\begin{align} \label{eq:zipprates}
	w_\pm (q) &=\Gamma \sqrt{g} \exp \left[ \pm \frac{f(q)}{2 k_B T} \mp \frac{\ln g}{2} \right],
\end{align}
with the function $f(q)$ given by Eq.~\eqref{eq:fx}. For $a=0$, at the phase transition point, $w_+(q)=w_-(q)=\Gamma \sqrt{g}$, and thus the drift term $\mu(q)$ disappears for all $q$. In other words, all $q$ are the attractors of the system dynamics. Consequently, the system does not converge to a stationary state with a well-defined $q$, corresponding to the minimum of the free energy density functional, but rather to a uniform probability distribution. Due to vanishing of the drift term, the relaxation to the uniform state is purely diffusive. Consequently, the slowest relaxation timescale of the system, which is related to the Liouvillian gap, is determined by the diffusion term, which scales as $N^{-2}$. This scaling corresponds to our analytic result~\eqref{eq:gapapproxa0} for the Liouvillian gap.  

We note that here we tacitly neglected higher-order terms of the Kramers-Moyal expansion (of order $O(N^3)$). This is justified as the drift term vanishes, and thus the diffusion term is dominant in the system dynamics. However, in a generic case where the drift term does not vanish, the higher-order terms become important when considering rare fluctuations far from the system attractors~\cite{hanggi1988bistability, gaveau1997master, vellela2009stochastic, hanggi1982stochastic} (among others, their neglect violates the thermodynamic consistency~\cite{FalascoReview, gopal2022large}). In particular, they strongly influence the transition rates between attractors considered in the next paragraph~\cite{kessler2007extinction}. We further note that similar arguments based on diffusive mechanism have been used to explain the critical slowing down at continuous phase transitions, where the slowest relaxation scale also scales as a power law with the system size~\cite{vankampen1981stochastic,gardiner2009stochastic}, or the power-law scaling of the Liouvillian gap in systems with periodic attractors~\cite{gaspard2002trace,gaspard2002correlation,gonze2002biochemical,remlein2022coherence}. Although in those cases the drift term is nonvanishing (in contrast to our example), the slowest relaxation timescale is determined by dynamics close to the system attractors, where the higher-order terms beyond the diffusive one can be neglected.

\subsubsection{Large $a$: exponential scaling}
We now focus on the opposite regime where $a$ is relatively large, and thus the temperature window of the phase coexistence is wide. In such a case, the Liouvillian gap is determined by the slowest process, namely the stochastic switching between the coexisting closed and open attractors. As shown in Refs.~\cite{hanggi1984bistable,hinch2005exponentially}, in the large $N$ limit, the Liouvillian gap is then asymptotically given by an Arrhenius-like formula
\begin{align} \label{arrh}
	\lambda \asymp A \exp(-N \beta_c \Delta \mathcal{F}),
\end{align}
where $A$ is the pre-exponential factor, dependent on system kinetics, and
\begin{align} \label{eq:deltafgen}
	\Delta \mathcal{F} \equiv \mathcal{F}(q_\text{max}) -\mathcal{F}(q_\text{eq}) \quad \text{for} \quad T=T_c
\end{align}
is the free energy density barrier separating the closed and open attractors. Here, $q_\text{eq}$ given by  Eq.~\eqref{eq:qeq} corresponds to the minimum of the free energy density functional, while
\begin{align} \label{eq:qmax}
q_\text{max} \equiv \operatorname*{arg\,max}_{q \in [0,1]} \mathcal{F}(q)
\end{align}
corresponds to its maximum. In our case, we have two degenerate minima with $q_\text{eq} \in \{0,1 \}$, and $q_\text{max}=1/2$. Using Eq.~\eqref{eq:freeendenzippspec}, the free energy density barrier reads
\begin{align} \label{eq:deltafspec}
	\Delta \mathcal{F} = \beta_c^{-1} \frac{a \ln g}{4(1+a)}.
\end{align}
The pre-exponential factor $A$ can be approximated analytically using the theory presented in Ref.~\cite{hinch2005exponentially} as 
\begin{align} \label{arrhamp}
	A \approx \frac{2 \Gamma \sqrt{g}}{\sqrt{\pi N}} \left( \frac{a \ln g}{1+a} \right)^{3/2}.
\end{align}
For details of the derivation, see Appendix~\ref{app:derexpgap}.

\begin{figure}
	\centering
	\includegraphics[width=0.9\linewidth]{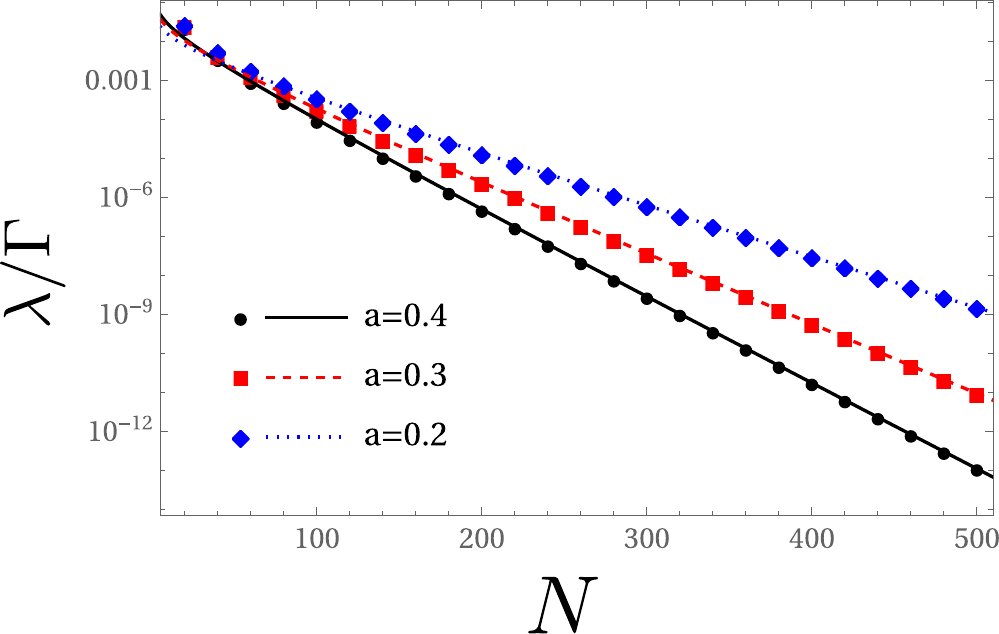}
	\caption{The Liouvillian gap, plotted in the logarithmic scale, as a function of $N$ for different values of $a$ calculated numerically (dots) and approximated using Eq.~\eqref{arrh} (lines). Parameters: $g=2$.}
	\label{fig:gapzipperlargea}
\end{figure}
As implied by Eq.~\eqref{arrh}, the Liouvillian gap closes exponentially with the system size\footnote{We also note that the pre-exponential factor, being proportional to $1/\sqrt{N}$, brings a subexponential correction to the exponential scaling. Arrhenius law with subexponential correction is known as the Eyring--Kramers formula~\cite{eyring1935activated, kramers1940brownian}.}. To illustrate that, Fig.~\ref{fig:gapzipperlargea} shows the scaling of the numerically calculated Liouvillian gap for different values of $a$, which is compared with the asymptotic formula~\eqref{arrh} with $A$ given by Eq.~\eqref{arrhamp}. The agreement for large system sizes is very good, which confirms the exponential character of the Liouvillian gap scaling in the considered regime.

\subsubsection{Small $a$: crossover from power-law to exponential scaling}
\begin{figure}
	\centering
	\includegraphics[width=0.9\linewidth]{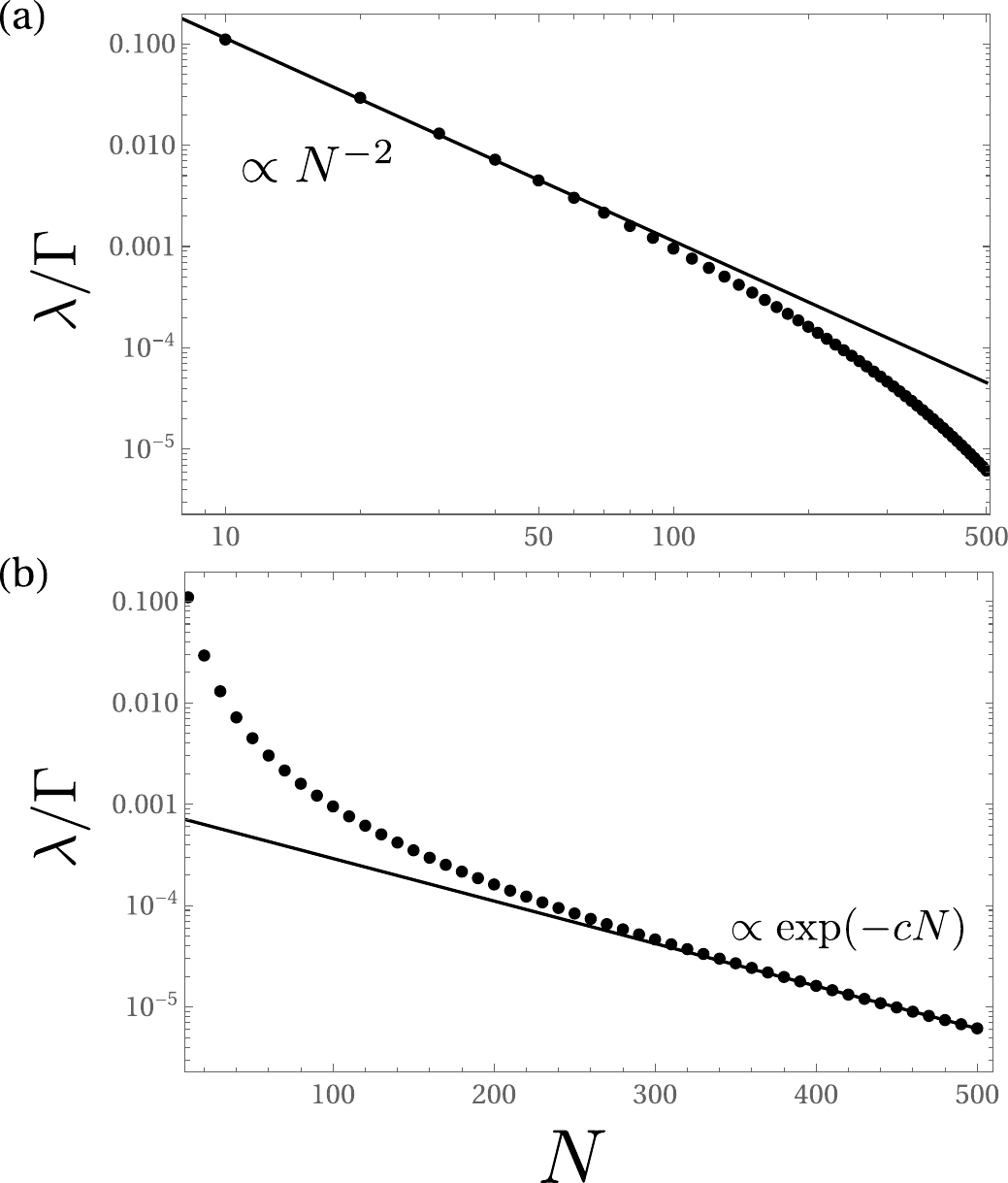}
	\caption{The Liouvillian gap as a function of $N$ for a relatively small $a=0.05$ plotted in the log-log scale (a) and the logarithmic scale (b). Black solid lines in (a) and (b) represent the power-law and exponential fit for small and large values of $N$, respectively. Exponential scaling is characterized by the factor $c \approx 0.0097$. Parameters: $g=2$.}
	\label{fig:gapzippersmalla}
\end{figure}
As discussed above, for $a=0$ the Liouvillian gap exhibits a power-law scaling, related to the diffusive nature of system relaxation to the stationary state, while for large $a$ it exhibits an exponential scaling, related to stochastic switching between attractors. The natural expectation is that, by increasing $a$ from 0 to finite values, one should continuously interpolate between those regimes. To illustrate that, in Fig.~\ref{fig:gapzippersmalla} we plot the scaling of the Liouvillian gap for a relatively small $a=0.05$ in the log-log scale (a) and the logarithmic scale (b). As may be observed, for small $N$ the system exhibits a power-law behavior, with the Liouvillian gap scaling approximately as $N^{-2}$, as in the case of $a=0$. This is represented by the black solid line in Fig.~\ref{fig:gapzippersmalla}~(a). On the other hand, for large $N$ it starts to exhibit an exponential scaling, which is represented by the black solid line in Fig.~\ref{fig:gapzippersmalla}~(b). Thus, one observes a crossover from power-law to exponential scaling with increasing $N$. As implied by comparison with Fig.~\eqref{fig:gapzipperlargea}, the onset of exponential behavior shifts toward the higher values of $N$ as $a$ decreases.

We can qualitatively relate the crossover behavior to the fact that the free energy density barrier $\Delta \mathcal{F}$ is relatively low for small values of $a$, and thus the switching between closed and open states does not determine the slowest timescale of the system for small values of $N$. Rather, as for $a=0$, the diffusive mechanism appears to be dominant. Consequently, for small $\Delta \mathcal{F}$, the onset of exponential behavior can shift toward very large values of $N$. This may hinder the observability of exponential scaling when the system sizes accessible in simulations or experiments are limited. 

\subsection{Fluctuations scaling}
As discussed in Sec.~\ref{sec:spectheorfluct}, eigenvalues of the Liouvillian, including the Liouvillian gap, are strictly connected to the properties of dynamical fluctuations of observables. In this subsection, we investigate whether the scaling of the Liouvillian gap manifests itself in the corresponding scaling of fluctuations. Specifically, we analyze the zero-frequency noise of the number of closed links. It can be calculated using Eq.~\eqref{zerofreqnoise}, with the observable $\bar{o}$ being the number operator defined as
\begin{align}
	\bar{n} \equiv \text{diag} \left(0,1,2,\ldots,N \right).
\end{align}

\subsubsection{$a=0$}
We first focus on the case of $a=0$, where the spectrum of the rate matrix can be found analytically. To apply the methods from Sec.~\ref{sec:spectheorfluct}, we use a quantum-like bra-ket notation with $\mathbf{p}_{ss} \equiv |\rho_{ss} \rrangle$. The steady state and the corresponding left eigenvector read as
\begin{align}
|\rho_{ss} \rrangle&=(1,\ldots,1)^T/(N+1), \\
\llangle \mathds{1}|&=(1,\ldots,1).
\end{align}
The other right eigenvectors of the rate matrix (for $j>0$) read~\cite{britanak2010discrete}
\begin{align}
| x_j \rrangle=\left( x^{(j)}_0,\ldots,x^{(j)}_N \right)^T,
\end{align}
where
\begin{align}
x_k^{(j)}=\sqrt{\frac{2}{N+1}} \cos \left[\frac{\pi j (k+\tfrac{1}{2})}{N+1} \right].
\end{align}
They correspond to elements of DCT-II matrix defined in the theory of cosine and sine transforms. The left eigenvectors read simply $\llangle y_j |=(|x_j \rrangle)^T$. Using that, we calculate the overlap functions given by Eq.~\eqref{eq:overlap} as
\begin{align}
\mathcal{A}^{(j)}_{nn}=\frac{\left[1-(-1)^j\right] \sin ^2\left(\frac{\pi  j}{N+1}\right) \csc ^6\left[\frac{\pi  j}{2 (N+1)}\right]}{16 (N+1)^2}.
\end{align}
For $j \ll N$, they can be approximated using the Taylor expansion as
\begin{align}
\mathcal{A}^{(j)}_{nn} \approx \frac{4 \left[1-(-1)^j\right]N^2}{\pi^4 j^4}.
\end{align}
Using the spectral decomposition of noise [Eq.~\eqref{zerofreqnoise}], it can be observed that the noise components $-\mathcal{A}^{(j)}_{nn}/\lambda_j$ decrease very quickly with $j$. Thus, at the first level of approximation, noise is given by the component related to the Liouvillian gap:
\begin{align} \label{eq:fluctapproxa0}
	S_{nn}(0) \approx \frac{2\mathcal{A}^{(1)}_{nn}}{\lambda} \approx \frac{16N^4}{\pi^6 \Gamma \sqrt{g}}
\end{align}

\begin{figure}
	\centering
	\includegraphics[width=0.9\linewidth]{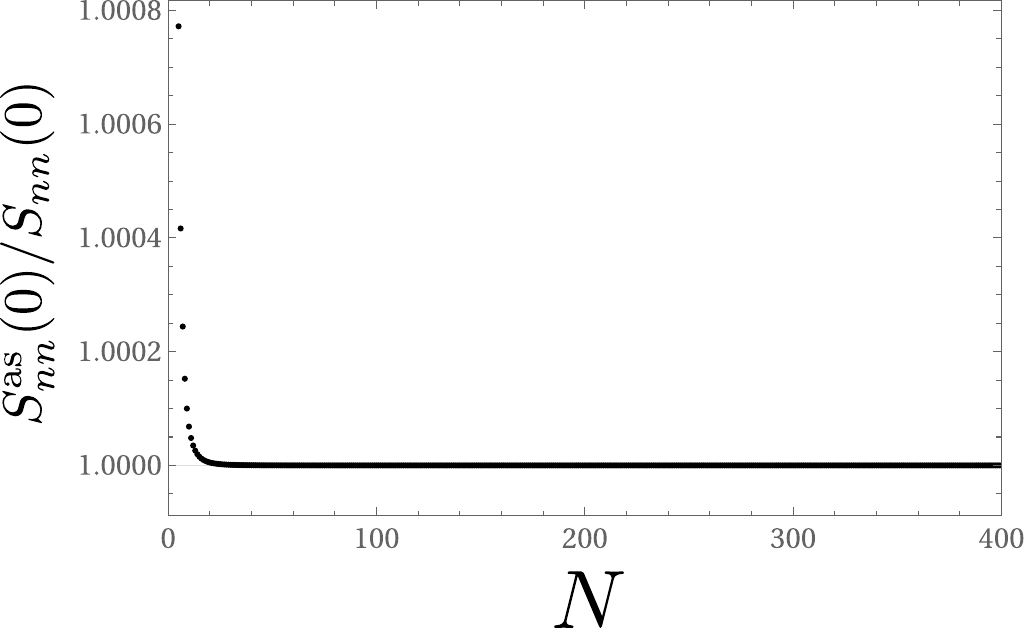}
	\caption{The ratio of the asymptotic noise formula $S_{nn}^\text{as}(0)$, given by Eq.~\eqref{eq:fluctasympa0}, and the exactly calculated noise $S_{nn}(0)$ for the molecular zipper with $a=0$.}
	\label{fig:fluctapproxa0}
\end{figure}
A more refined asymptotic formula for $S_{nn}(0)$ is obtained by calculating the limit
\begin{align}
S_{nn}(0) \asymp -N^4 \lim_{N \rightarrow \infty} \frac{1}{N^4} \sum_{j=1}^N \frac{2 \mathcal{A}^{(j)}_{nn}}{\lambda_j}.
\end{align}
This yields
\begin{align} \label{eq:fluctasympa0}
	S_{nn}(0) \asymp S^\text{as}_{nn}(0) =\frac{N^4}{60 \Gamma \sqrt{g}},
\end{align}
which is just about 1.0014 times larger than the approximate formula~\eqref{eq:fluctapproxa0}. Thus, analogously to the Liouvillian gap, fluctuations exhibit a power-law scaling. Fig.~\ref{fig:fluctapproxa0} presents the ratio of this asymptotic power-law formula and the exact result as a function of $N$. We can observe that the noise follows the asymptotic $\propto N^4$ behavior very well already for  small system sizes (with the relative error below 0.1\%), and converges very fast to the asymptotic formula.

\subsubsection{Large $a$} \label{subsec:fluctzipplargea}
Let us now consider the opposite case of large $a$. In this situation, the noise is mostly determined by the slowest process, namely, the switching between the closed and open attractors. This switching can be described as a Markovian process described by an effective rate matrix $W_\text{eff}$ acting on the probability vector $\mathbf{p}_\text{eff} \equiv (p_O,p_C)^T$, where $p_O$ ($p_C$) is the probability of the open (closed) attractor. Using the theory presented in Ref.~\cite{hinch2005exponentially}, the effective rate matrix can be approximated as
\begin{align}
W_\text{eff} \approx \frac{\lambda}{\mathcal{T}_++\mathcal{T}_-} \begin{pmatrix} -\mathcal{T}_- & \mathcal{T}_+ \\ \mathcal{T}_- & -\mathcal{T}_+ \end{pmatrix},
\end{align}
whose Liouvillian gap is equal to $\lambda$. The parameters $\mathcal{T}_\pm$ are defined in Eq.~\eqref{eq:tpm} in Appendix~\ref{app:derexpgap}; as shown there, for the case considered $\mathcal{T}_+=\mathcal{T}_-$. Following Ref.~\cite{nguyen2020exponential}, fluctuations can then be characterized using an effective two-state model. This is done using the approach from Sec.~\ref{sec:spectheorfluct}, with the total rate matrix replaced by the effective matrix $W_\text{eff}$, and using the effective number operator $\bar{n}_\text{eff} \equiv \text{diag}(0,N)$. Noise can then be approximated as
\begin{align} \label{eq:fluctlargea}
	S_{nn}(0) \approx \frac{N^2}{2 \lambda}.
\end{align}
Since the Liouvillian gap closes exponentially, one may expect an exponential divergence of fluctuations.

\begin{figure}
	\centering
	\includegraphics[width=0.9\linewidth]{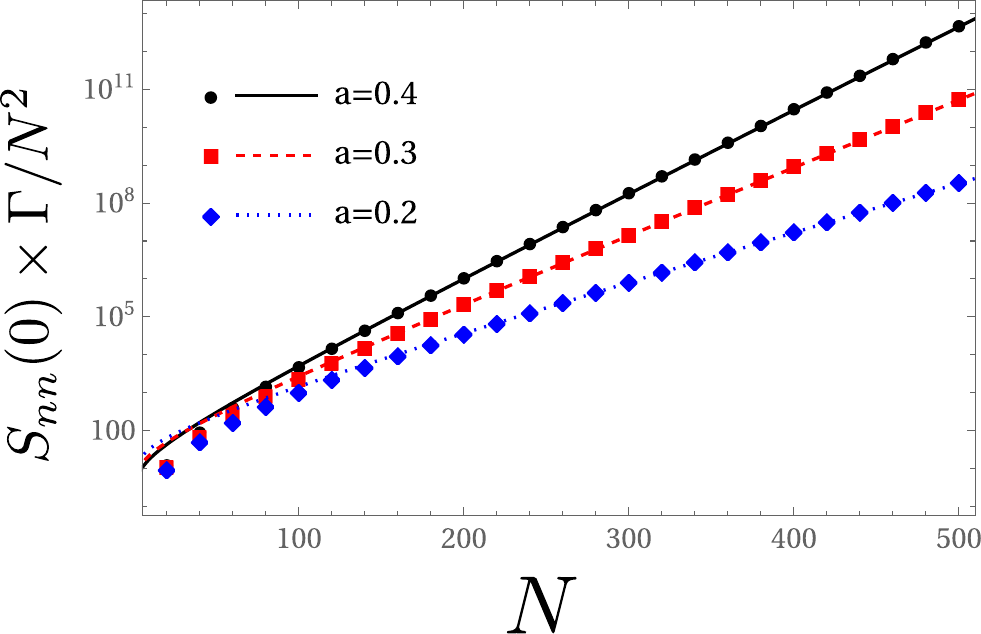}
	\caption{The zero frequency noise $S_{nn}(0)$ as a function of $N$ for different values of $a$ calculated numerically (dots) and approximated using Eq.~\eqref{eq:fluctlargea} (lines), plotted in the logarithmic scale. Parameters: $g=2$.}
	\label{fig:fluctzipperlargea}
\end{figure}
In Fig.~\ref{fig:fluctzipperlargea} we present the scaling of the numerically calculated zero-frequency noise with $N$ for different values of $a$ (dots), compared with the approximate formula~\eqref{eq:fluctlargea}, with the Liouvillian gap evaluated using Eq.~\eqref{arrh} (lines). As in the case of the Liouvillian gap, we observe a very good agreement, which confirms the exponential scaling of fluctuations in the regime considered.

\subsubsection{Small $a$}
\begin{figure}
	\centering
	\includegraphics[width=0.9\linewidth]{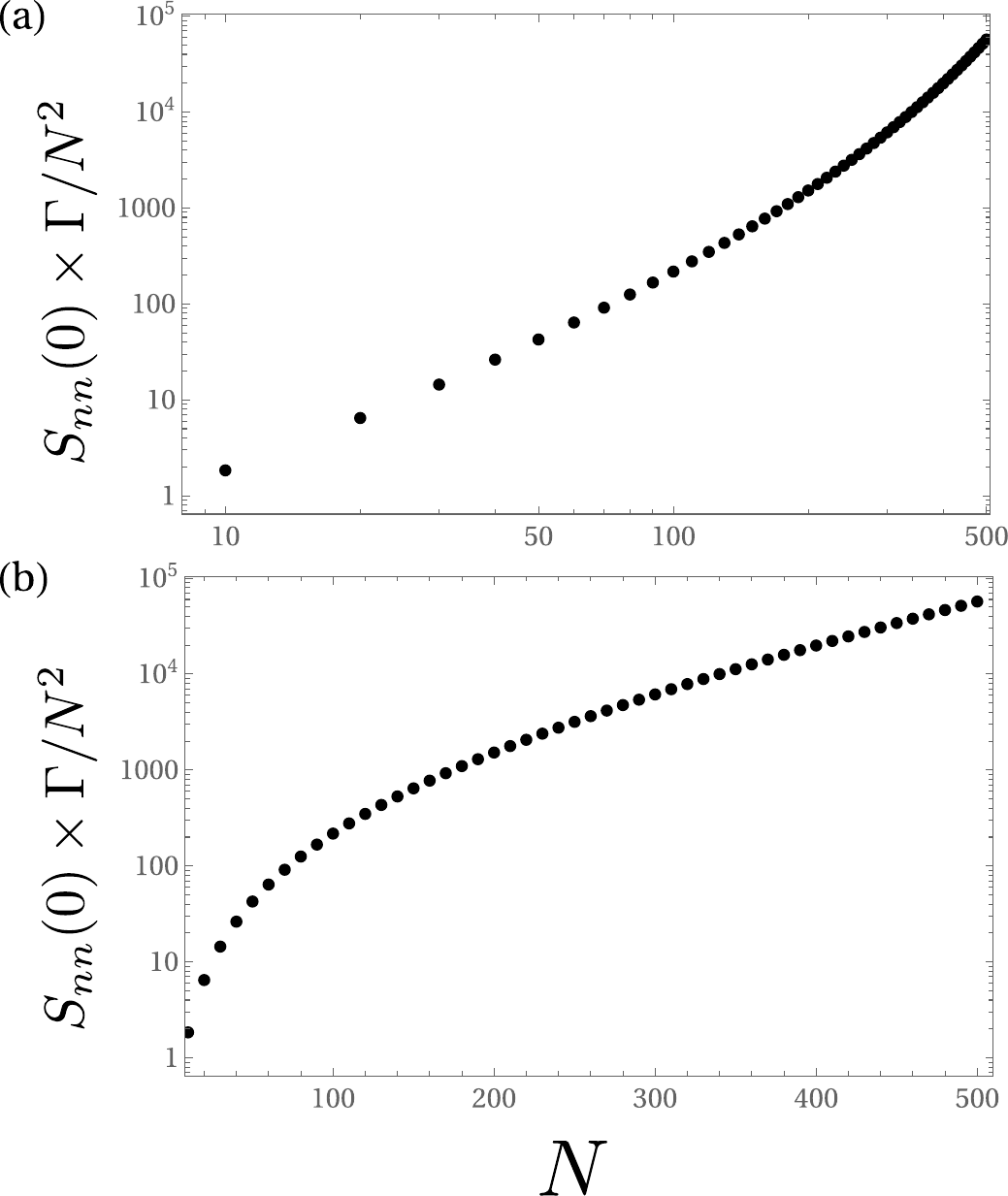}
	\caption{The zero frequency noise $S_{nn}(0)$ as a function of $N$ for a relatively small $a=0.05$ plotted in the log-log scale (a) and the logarithmic scale (b). Parameters: $g=2$.}
	\label{fig:fluctsmalla}
\end{figure}
Finally, let us consider the regime where the parameter $a$ is relatively small (here $a=0.05)$. The scaling of fluctuations in such a case is presented in Fig.~\ref{fig:fluctsmalla}. Let us just briefly comment that we can see the same crossover from power-law to exponential behavior as for the Liouvillian gap, with the exponential scaling becoming clearly apparent only above a certain threshold $N \approx 200$.

\section{Open quantum Ising model} \label{sec:ising}
\subsection{Description of the model}
To illustrate the generality of our observations, we now turn to a nonequilibrium quantum scenario. Specifically, we focus on the open quantum Ising model~\cite{wang2021dissipative} (a special instance of the dissipative LMG model~\cite{morrison2008collective, lee2013unconventional, lee2014dissipative, ferreira2019lipkin, debecker2023spectral, song2023crossover}) described by the master equation
\begin{align}
\dot{\rho}=-i[H,\rho]+\sum_{\pm} \frac{\gamma_\pm}{N} \mathcal{D}[\hat{S}_\pm](\rho)
\end{align}
with the effective Hamiltonian
\begin{align}
H \equiv-\frac{J}{N} \hat{S}_x^2+h\hat{S}_z.
\end{align}
Here, $\hat{S}_{x,y,z}$ are spin-$N/2$ operators, $J \geq 0$ is the Ising-type ferromagnetic coupling, $h$ is the magnetic field, $\hat{S}_{\pm}=\hat{S}_x \pm i \hat{S}_y$ are spin raising and lowering operators, and $\gamma_{\pm}$ are the associated dissipation rates. This model can be realized using an ensemble of $N$ spins-$1/2$ undergoing a collective dissipation, which preserves the total angular momentum. We note that here we generalize the original model where $\gamma_+=0$. We note further that in Ref.~\cite{wang2021dissipative} the Liouvillian was expressed in terms of Pauli matrices rather than spin operators (which differ by a factor of 2), and thus the expressions presented there may differ by some scale factors.

The nonequilibrium phase transitions in the system have been thoroughly studied in Ref.~\cite{wang2021dissipative}. Let us just review some of the most important aspects. In the large $N$ limit, the system dynamics can be described using mean field equations for normalized magnetization components $m_{x,y,z} \equiv 2 \langle \hat{S}_{x,y,z} \rangle/N$. They read
\begin{align} \nonumber
\dot{m}_x&=-h m_y+\gamma m_x m_z/2, \\ \label{eq:ising-meanf}
\dot{m}_y &=J m_x m_z +h m_x+\gamma m_y m_z/2, \\ \nonumber
\dot{m}_z &=-J m_x m_y-\gamma (m_x^2+m_y^2)/2.
\end{align}
Here we used the effective dissipation rate
\begin{align} \label{eq:effdisrate}
\gamma \equiv \gamma_--\gamma_+,
\end{align}
which (in contrast to the original model with $\gamma_+=0$) can be either positive or negative. The equations preserve the total magnetization $m=\sqrt{m_x^2+m_y^2+m_z^2}$; here we focus on the case of $m=1$. We further define the magnetization vector $\mathbf{m}=(m_x,m_y,m_z)$.

As discussed in Ref.~\cite{wang2021dissipative}, the mean field equations admit several stable and unstable fixed points. We focus only on the stable points. For simplicity, we restrict ourselves to the case of $h \geq 0$. Then, for $\gamma<0$, the system has a unique fixed point $\mathcal{N}$ with
\begin{align}
\mathbf{m}^\mathcal{N}=(0,0,1).
\end{align}
$\mathcal{N}$ corresponds therefore to the ``north pole'' of the Bloch sphere. For $\gamma>0$ this solution is unstable. The opposite ``south pole'' fixed point $\mathcal{S}$ is associated with the vector
\begin{align} \label{eq:southpole}
\mathbf{m}^\mathcal{S}=(0,0,-1).
\end{align}
This solution is stable for $\gamma>0$ except for $0 \leq \gamma \leq 1$ and $B_- \leq h \leq B_+$, where
\begin{align}
B_\pm=\frac{J}{2} \left(1 \pm \sqrt{1-\gamma^2/J^2} \right).
\end{align}
Additionally, there are two other fixed points $\mathcal{P}_\pm$ with finite magnetization components in the $x$ and $y$ directions. Their components read
\begin{align} \nonumber
m_x^{\mathcal{P}_\pm} &= \pm \sqrt{\left(B_+-\frac{4h^2}{\gamma^2} B_-\right)/J}, \\
m_y^{\mathcal{P}_\pm} &=\frac{\gamma}{2h} m_x^{\mathcal{P}_\pm} m_z^{\mathcal{P}_\pm}, \\ \nonumber
m_z^{\mathcal{P}_\pm} &=-\frac{h}{B_+}.
\end{align}
\begin{figure}
	\centering
	\includegraphics[width=0.9\linewidth]{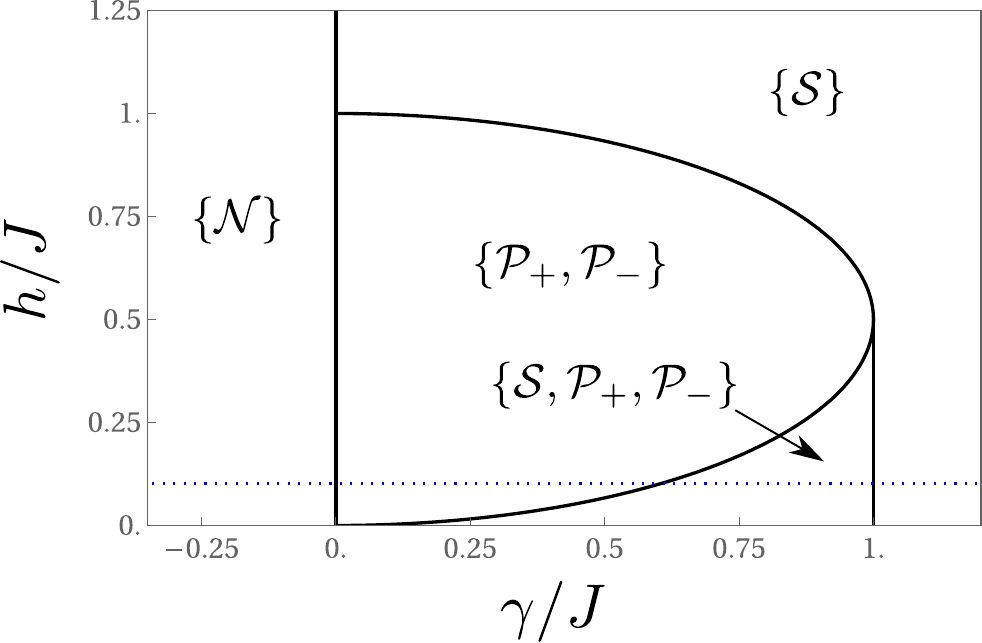}
	\caption{Phase diagram representing the stable fixed points of mean field equations for the open quantum Ising model. The symbols within the curly braces denote the fixed points that are stable in a given region. The horizontal blue dotted line corresponds to $h=0.1J$ used in the magnetization plot in Fig.~\ref{fig:magnising}.}
	\label{fig:phasediagising}
\end{figure}
These solutions are stable for $0 \leq \gamma \leq 1$ and $h \leq B_+$. The phase diagram representing the stable fixed points is presented in Fig.~\ref{fig:phasediagising}. The symbols within the curly braces denote the fixed points that are stable in a given region. In particular, we note that for $0 <\gamma <J$ and $h<B_+$ there are three stable fixed points: $\mathcal{S}$, $\mathcal{P}_+$ and $\mathcal{P}_-$.

\begin{figure}
	\centering
	\includegraphics[width=0.9\linewidth]{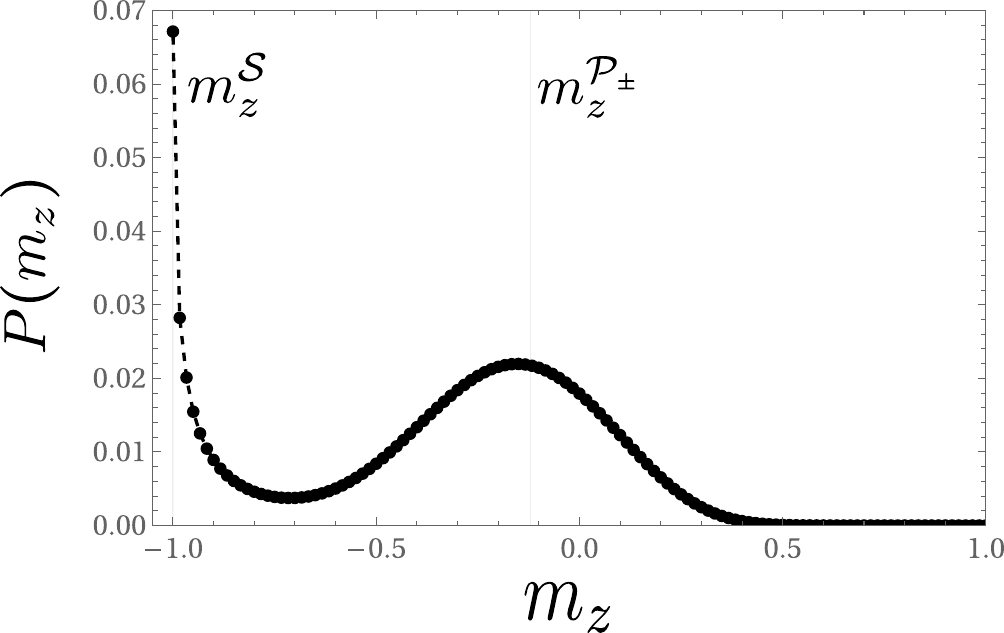}
	\caption{The probability distribution of the normalized magnetization $m_z \equiv 2\langle \hat{S}_z \rangle/N$ in the phase coexistence region for $h=0.1J$, $\gamma_+=0.25J$, $\gamma_-=J$, and $N=120$. The vertical solid lines represent the mean field solutions $m_z^\mathcal{S}=-1$ and $m_z^{\mathcal{P}_\pm} \approx -0.12$. The dashed line is added for eye guidance.}
	\label{fig:bimodal}
\end{figure}
The mean field approach (which corresponds to taking the thermodynamic limit $N \rightarrow \infty$ before the long time limit $t \rightarrow \infty$) admits the presence of multiple stable fixed points. In contrast, the master equation predicts the presence of a unique stationary state, corresponding to a definite value of expected values of observables. This apparent discrepancy, known as the Keizer's paradox~\cite{keizer1978thermodynamics}, can be resolved by noting that when the mean field equations admit three fixed points $\mathcal{S}$, $\mathcal{P}_+$, and $\mathcal{P}_-$, the probability distribution of the normalized magnetization $m_z$ has a bimodal structure, with the peaks of the probability distribution approximately corresponding to the mean field solutions (Fig.~\ref{fig:bimodal}). For large system sizes, this leads to the double-minimum structure of the rate function (see Sec.~\ref{sec:disc}), and the attractors of the mean field dynamics correspond to the minima of the rate function~\cite{hanggi1982stochastic,FalascoReview}. As discussed before, the lifetimes of metastable attractors become infinite in the thermodynamic limit, which corresponds to the presence of multiple stable fixed points of the mean field equations.

\begin{figure}
	\centering
	\includegraphics[width=0.9\linewidth]{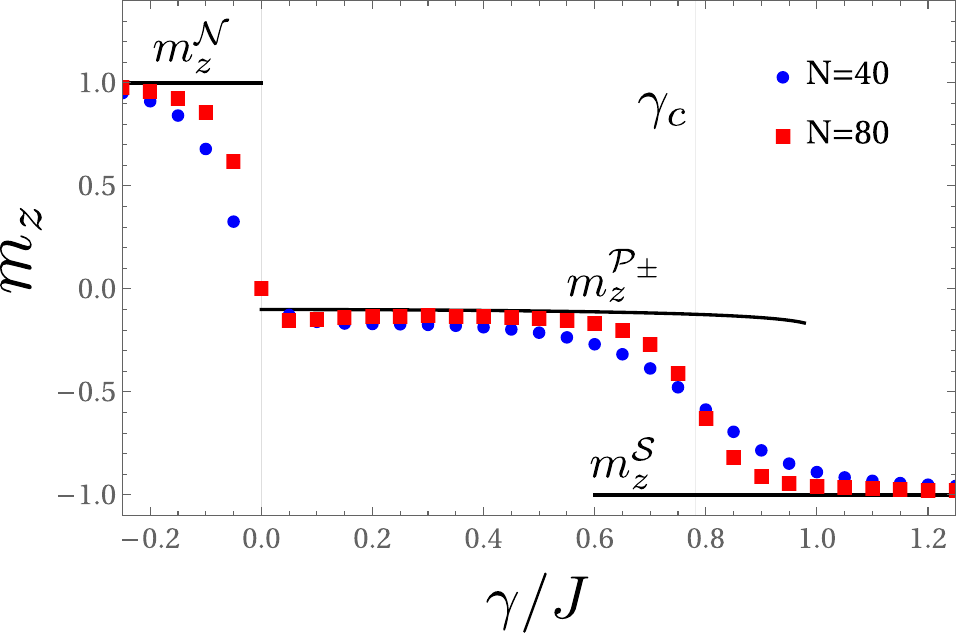}
	\caption{The normalized magnetization $m_z \equiv 2\langle \hat{S}_z \rangle/N$ as a function of $\gamma$ for $h=0.1J$ (denoted by blue dotted line in Fig.~\ref{fig:phasediagising}). The black solid lines denote the fixed points of mean field equations, while the dots denote the numerical results for finite $N$. $\gamma_c \approx 0.782 J$ denotes the position of the second discontinuous phase transition. Parameters: $\gamma_+=0.25J$.}
	\label{fig:magnising}
\end{figure}
Let us now analyze the behavior of the normalized magnetization $m_z$ as a function of the parameter $\gamma$, for a fixed $h=0.1$. This is presented in Fig.~\ref{fig:magnising}. As shown, magnetization exhibits two rapid jumps at $\gamma=0$ and $\gamma=\gamma_c \approx 0.782 J$ (the latter value has been estimated by finding the crossing point of magnetization curves for $N=60$ and $N=80$). These jumps become increasingly sharp with $N$ and thus, in the thermodynamic limit, they correspond to discontinuous phase transitions. The first transition occurs at the point where two opposite dissipation rates $\gamma_+$ and $\gamma_-$ compensate. At this point, the fixed point $\mathcal{N}$ becomes unstable at the expense of two fixed points $\mathcal{P}_\pm$, and thus the magnetization jumps from $m_z^\mathcal{N}$ to $m_z^{\mathcal{P}_\pm}$. Notably, this transition does not occur within a phase coexistence region: only the fixed point $\mathcal{N}$ is stable below $\gamma<0$, and only fixed points $\mathcal{P}_+$, and $\mathcal{P}_-$ are stable for $\gamma>0$. Furthermore, at the phase transition point itself, the effective dissipation rate $\gamma$ vanishes, and thus the system has an infinite number of stable periodic orbits corresponding to purely unitary dynamics. In the master equation picture, the density matrix of the system becomes fully mixed at this point. This is analogous to the molecular zipper model with $a=0$. 

The second transition occurs, instead, within the phase coexistence region where three fixed points of the mean field equations, $\mathcal{S}$, $\mathcal{P}_+$, and $\mathcal{P}_-$, are stable. At the phase transition point $\gamma_c$, the attractor $\mathcal{S}$ becomes absolutely stable (i.e., probabilistically dominantly occupied) at the expense of $\mathcal{P}_\pm$, and thus the stationary magnetization jumps from $m_z^{\mathcal{P}_\pm}$ to $m_z^{\mathcal{S}}$. This is analogous to the molecular zipper with $a>0$.

\subsection{Scaling of the Liouvillian gap and fluctuations}

\subsubsection{$\gamma=0$: power-law scaling}
%
\begin{figure}
	\centering
	\includegraphics[width=0.9\linewidth]{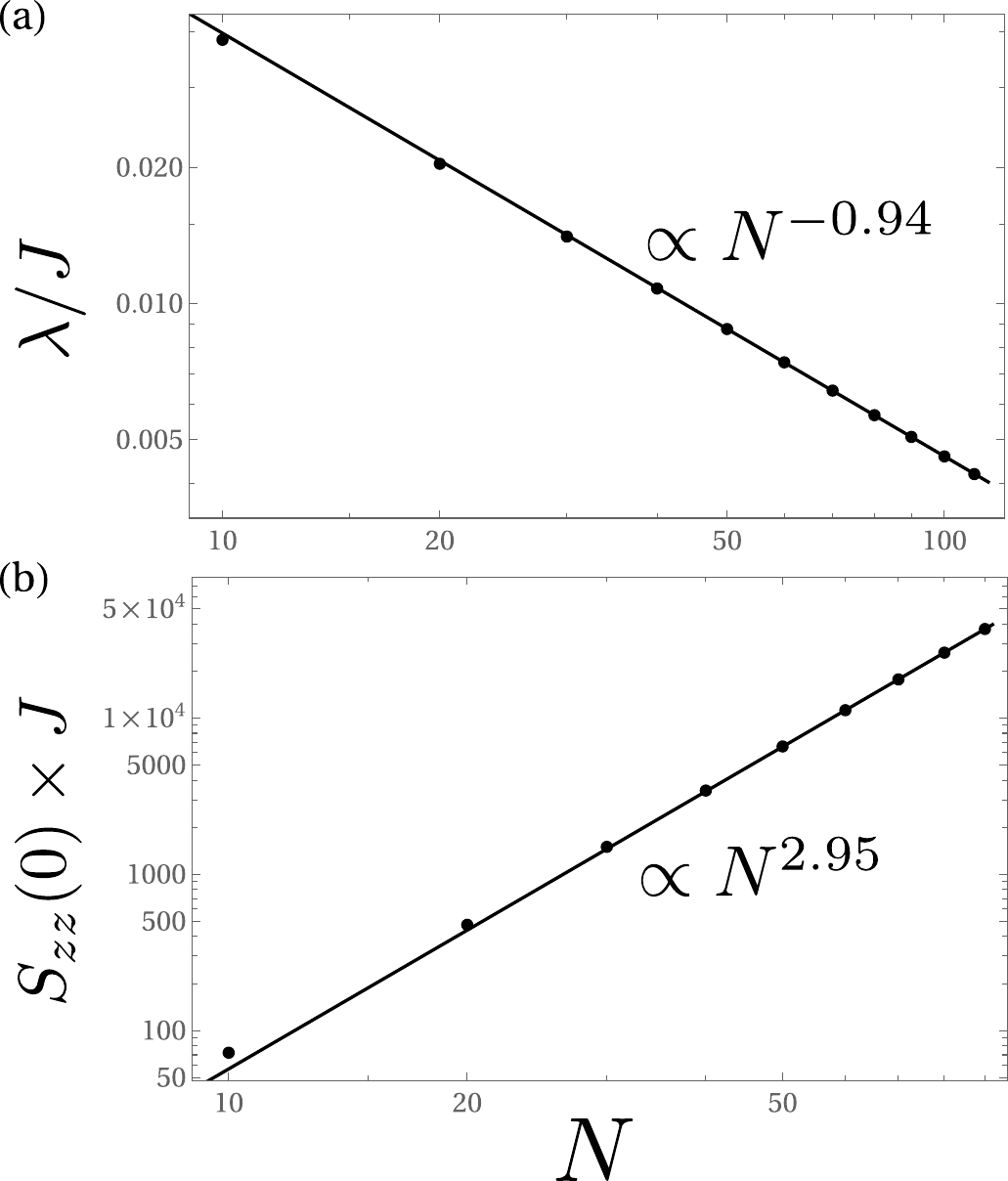}
	\caption{The finite-size scaling with $N$ of the Liouvillian gap (a) and fluctuations (b) for $\gamma=0$, plotted in the log-log scale. The results are denoted by dots, black solid lines correspond to power-law fits for large $N$. Parameters: $h=0.1J$, $\gamma_+=0.25J$.}
	\label{fig:gapfluctisingg0}
\end{figure}
Let us now consider the finite-size scaling of the Liouvillian gap and fluctuations at the mentioned phase transition points. Specifically, we analyze the zero-frequency noise of magnetization in the $z$ direction, denoted as $S_{zz}(0)$, which corresponds to taking $\hat{o}=\hat{S}_z$ in Eq.~\eqref{noisespec}. We first focus on the transition without phase coexistence, that occurs at $\gamma=0$. As shown in Fig.~\ref{fig:gapfluctisingg0}, the Liouvillian gap and fluctuations exhibit then a power-law scaling for large $N$. This is analogous to the molecular zipper with $a=0$, where the phase transition also occurs without phase coexistence. We further note that a power-law scaling has also been reported for a discontinuous phase transition without phase coexistence in a similar model~\cite{ferreira2019lipkin}. Thus, one may draw  quite general conclusion that when the discontinuous phase transition does not occur within a finite coexistence region, it is associated with a power-law scaling of the Liouvillian gap and fluctuations. However, we will add a certain reservation to this conclusion is Sec.~\ref{sec:sdm}.

We can rationalize the power-law scaling of the Liouvillian gap using a diffusive mechanism analogous to that used for the molecular zipper. To this end, we note that the system dynamics can be described by a partial differential equation for some quantum quasiprobability distribution $\chi(\boldsymbol{m})$, e.g., Glauber, Wigner or Husimi distribution~\cite{walls1978non,drummond1978volterra,carmichael1980analytical,smith1988phase,merkel2021phase,dubois2021semi}. Independent of the choice of distribution considered, such an equation has a general form resembling Kramers-Moyal expansion of the classical master equation,
\begin{align} &
\partial_t \chi(\boldsymbol{m})=-\sum_{i \in \{x,y,z \}} \frac{\partial}{\partial m_i} \left[\mu_i(\boldsymbol{m}) \chi(\boldsymbol{m}) \right] \\ \nonumber &+\sum_{i,j \in \{x,y,z \}} \frac{\partial}{\partial m_i} \frac{\partial}{\partial m_j} \left[D_{ij}(\boldsymbol{m}) \chi(\boldsymbol{m}) \right] +O(N^{-2}).
\end{align}
Here, $\boldsymbol{m}$ is the magnetization vector defined below Eq.~\eqref{eq:effdisrate}, $\mu_i(\boldsymbol{m})$ are the elements of the drift vector of order $O(1)$, related to the mean field equations~\eqref{eq:ising-meanf} as $\dot{m_i}=\mu_i(\boldsymbol{m})$, and $D_{ij}(\boldsymbol{m})$ are the elements of the diffusion matrix of order $O(1/N)$, whose form depends on the quasiprobability distribution considered. We recall that, at the phase transition point, the mean field equations have an infinite number of stable periodic orbits, while the master equation approach predicts relaxation to a fully mixed state. Therefore, in analogy to the molecular zipper, the system relaxation to the stationary state has a purely diffusive character, i.e., it corresponds to longitudinal diffusion along the periodic orbits and transverse diffusion between them~\cite{walls1978non,drummond1978volterra,carmichael1980analytical}. Thus, one may expect a $1/N$ scaling of the Liouvillian gap, corresponding to the scaling of the diffusion coefficient. The numerical results presented in Fig.~\ref{fig:gapfluctisingg0}~(a) show a scaling that is close but not exactly proportional to $1/N$; however, this discrepancy may be a result of the finite system sizes considered in the simulations.

\subsubsection{$\gamma=\gamma_c$: exponential scaling} \label{subsec:isingscalgc}
%
\begin{figure}
	\centering
	\includegraphics[width=0.9\linewidth]{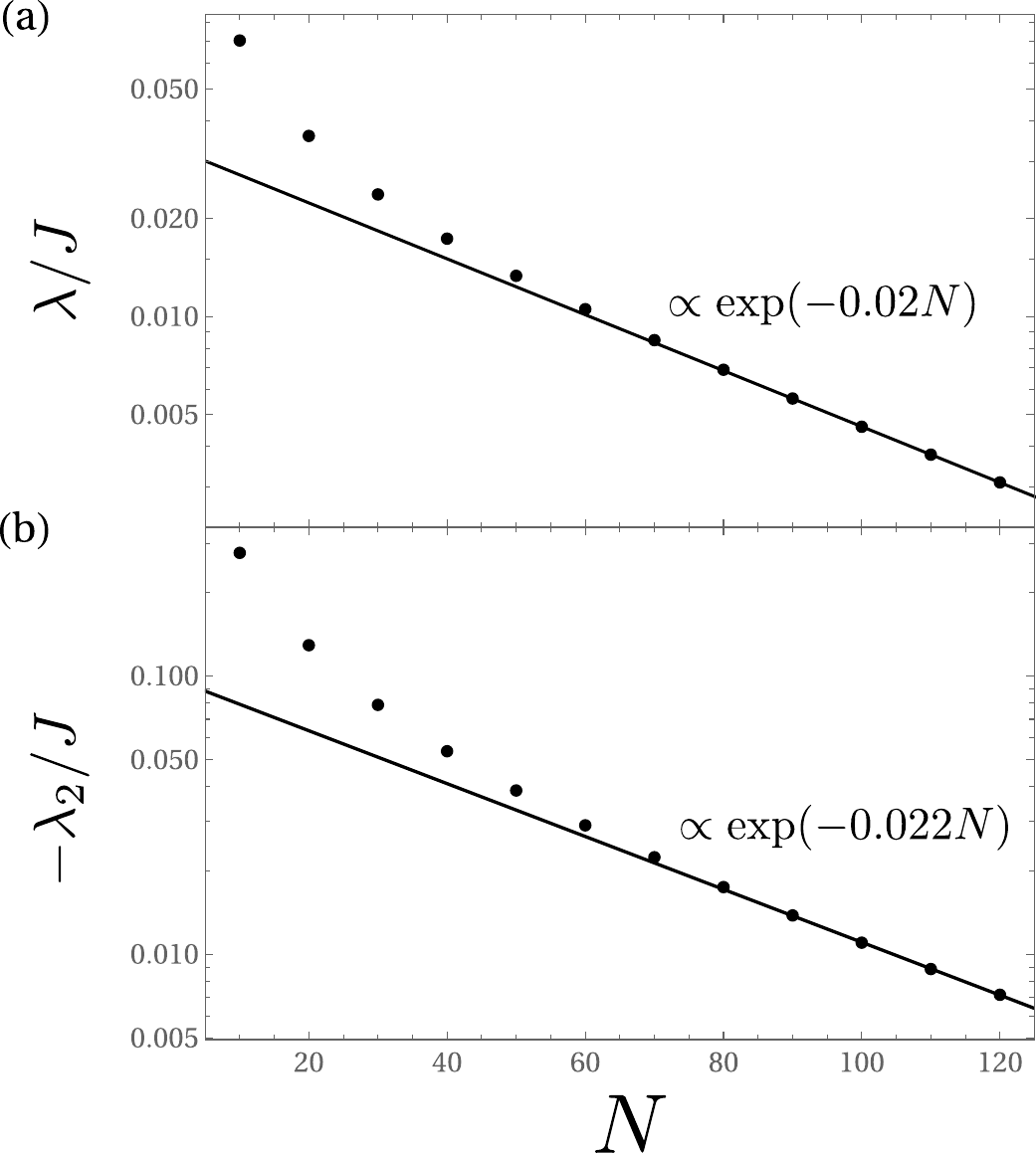}
	\caption{The finite-size scaling with $N$ of the Liouvillian gap $\lambda$ (a) and the eigenvalue $\lambda_2$ (b) for $\gamma=\gamma_c \approx 0.782J$, plotted in the logarithmic scale. The results are denoted by dots, black solid lines correspond to exponential fit for large $N$. Parameters: $h=0.1J$, $\gamma_+=0.25J$.}
	\label{fig:gapisinggc}
\end{figure}
Let us now turn to the transition with phase coexistence, that occurs at $\gamma=\gamma_c$. In this case, we observe that both the Liouvillian gap and the eigenvalue $\lambda_2$ asymptotically converge to 0 and exhibit a similar scaling behavior. This is related to the coexistence of three attractors, $\mathcal{S}$, $\mathcal{P}_+$, and $\mathcal{P}_-$, in the region around the phase transition point~\cite{macieszczak2021theory}. The finite-size scaling of the Liouvillian gap and $-\lambda_2$ is plotted in Fig.~\ref{fig:gapisinggc}. As in the case of molecular zipper, the scaling of both quantities exhibits a crossover behavior, becoming exponential for large $N$. This is consistent with the results from Ref.~\cite{wang2021dissipative} for the case of $\gamma_+=0$.

\begin{figure}
	\centering
	\includegraphics[width=0.85\linewidth]{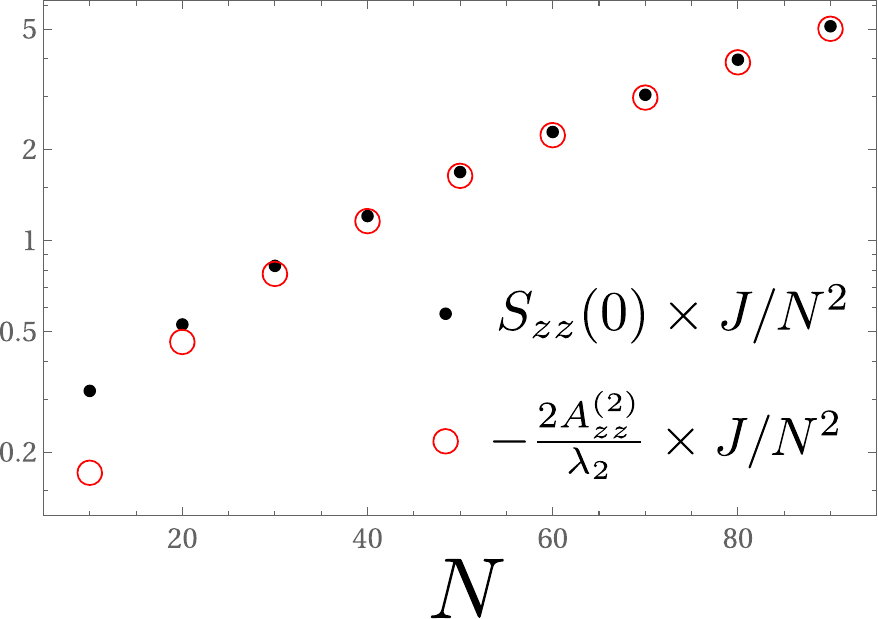}
	\caption{The finite-size scaling with $N$ of the zero-frequency noise $S_{zz}(0)$ (black dots) and the noise contribution $-2\mathcal{A}_{zz}^{(2)}/\lambda_2$ (red circles) for $\gamma=\gamma_c \approx 0.782J$, plotted in the logarithmic scale. Parameters: $h=0.1J$, $\gamma_+=0.25J$.}
	\label{fig:fluctisinggc}
\end{figure}
\begin{figure}
	\centering
	\includegraphics[width=0.85\linewidth]{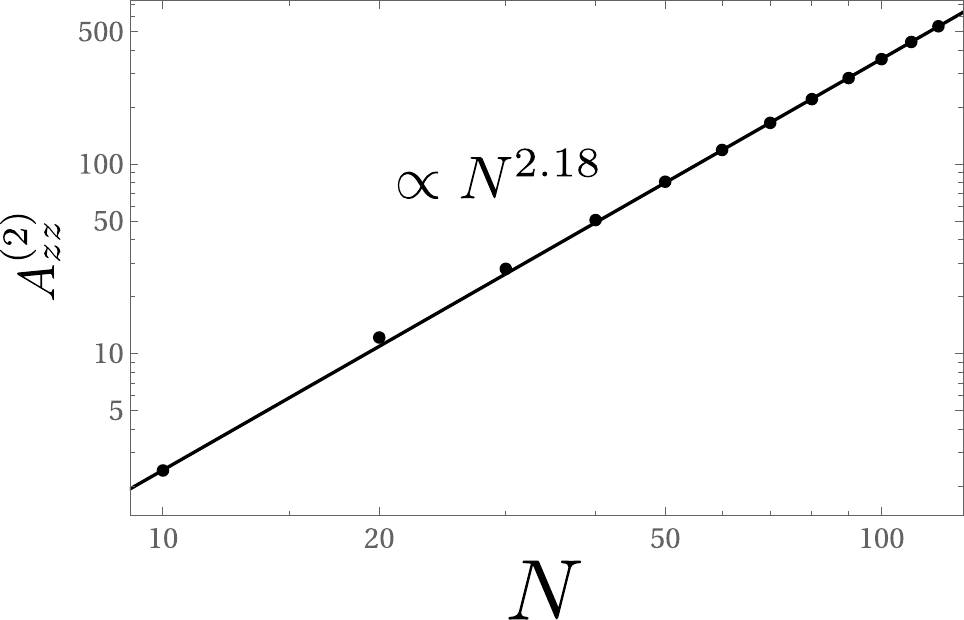}
	\caption{The finite-size scaling with $N$ of the overlap function $\mathcal{A}_{zz}^{(2)}$ for $\gamma=\gamma_c \approx 0.782J$, plotted in the log-log scale. The results are denoted by dots, the black solid line correspond to the power-law fit. Parameters: $h=0.1J$, $\gamma_+=0.25J$.}
	\label{fig:a2isinggc}
\end{figure}
In Fig.~\ref{fig:fluctisinggc} we present the finite-size scaling of the magnetization noise. It appears that for large $N$ it also exhibits an exponential scaling. However, this may not be fully clear for the range of $N$ we are able to simulate. To demonstrate it more convincingly, let us note that for large $N$ the noise becomes dominated by the contribution $-2\mathcal{A}_{zz}^{(2)}/\lambda_2$, associated with the eigenvalue $\lambda_2$. This contribution is plotted with red circles in Fig.~\ref{fig:fluctisinggc}. (Interestingly, the overlap function $\mathcal{A}_{zz}^{(1)}$, associated with the Liouvillian gap, vanishes. It may thus be inferred that this overlap function corresponds to the switching process between $\mathcal{P}_+$ and $\mathcal{P}_-$ attractors, which does not affect magnetization in the $z$ direction.) As shown in Fig.~\ref{fig:a2isinggc}, the overlap function $\mathcal{A}_{zz}^{(2)}$ exhibits a power-law scaling. (We note, however, that the estimated power-law behavior differs from the quadratic scaling, which can be predicted based on the two-state model from Sec.~\ref{subsec:fluctzipplargea}. We are unsure whether this is just a finite-size effect or this scaling is sustained for larger $N$.) Thus, one may infer that for large $N$ the fluctuations scaling becomes dominated by the exponential scaling of $\lambda_2$.

\section{Transition to time-dependent attractors : anisotropic LMG model} \label{sec:lmg}
Thus far, we analyzed the situations in which the mean field equations exhibit only fixed-point attractors. We now consider the case where the system undergoes a discontinuous phase transition without phase coexistence between the phases with fixed-point and time-dependent mean field attractors. This is the anisotropic LMG model with collective decay, described by the master equation~\cite{lee2014dissipative,debecker2023spectral}
\begin{align}
\dot{\rho}=-i[H,\rho]+\frac{\gamma}{N} \mathcal{D}[\hat{S}_-](\rho)
\end{align}
with the effective Hamiltonian
\begin{align}
H \equiv\frac{J}{N} \left( \hat{S}_x^2-\hat{S}_y^2 \right).
\end{align}
Here, as previously, $\hat{S}_{x,y,z}$ are spin-$N/2$ operators, $J$ is the interaction energy, and $\gamma$ is the dissipation rate. In the thermodynamic limit, the system is described by mean field equations
\begin{align} \nonumber
\dot{m}_x&=-J m_y m_z+\gamma m_x m_z/2, \\
\dot{m}_y &=-J m_x m_z +\gamma m_y m_z/2, \\ \nonumber
\dot{m}_z &=2J m_x m_y-\gamma (m_x^2+m_y^2)/2.
\end{align}
\begin{figure}
	\centering
	\includegraphics[width=0.9\linewidth]{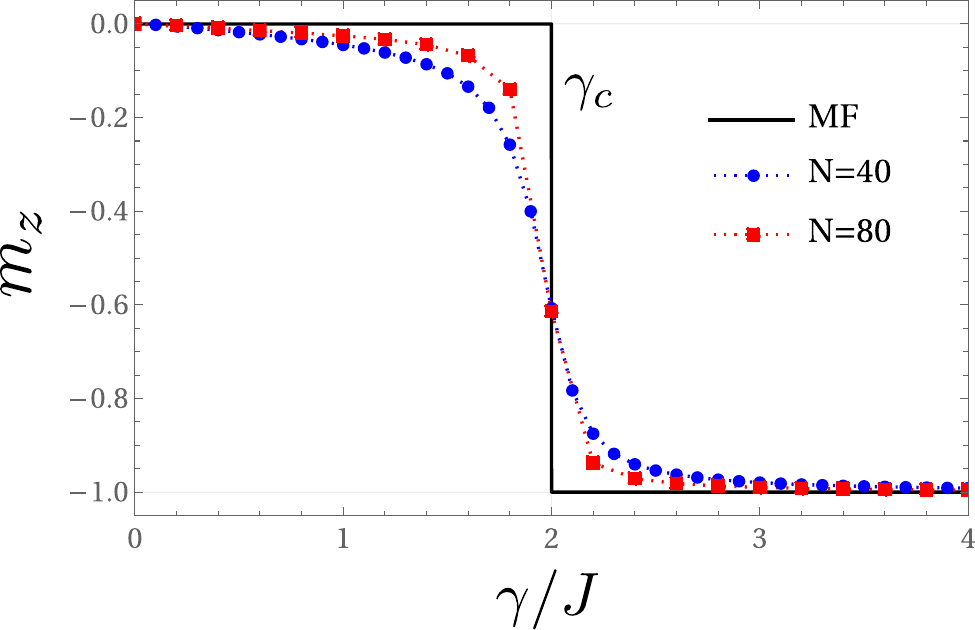}
	\caption{The normalized magnetization $m_z \equiv 2\langle \hat{S}_z \rangle/N$ as a function of $\gamma$. The black solid line are the mean field (MF) predictions for the time-averaged magnetization, while the dots represent the master equation results for finite system sizes $N$.}
	\label{fig:magnlmg}
\end{figure}
For $\gamma>\gamma_c=2J$, the system has a unique ``south pole'' fixed point $\boldsymbol{m}^\mathcal{S}=(0,0,-1)$ (see Eq.~\eqref{eq:southpole}). Instead, for $\gamma \leq \gamma_c$, the mean field equations do not have a unique attractor, but rather infinitely many periodic orbits corresponding to different initial conditions; see Ref.~\cite{lee2014dissipative} for a detailed analysis. For each orbit, the time-averaged magnetization in the z-direction $\overline{m_z(t)}=0$. In the master equation picture, this corresponds to a discontinuous phase transition at $\gamma=\gamma_c$, with the average magnetization changing from $m_z=0$ for $\gamma<\gamma_c$ to $m_z=-1$ for $\gamma>\gamma_c$. This is illustrated by finite-size results in Fig.~\ref{fig:magnlmg}, which asymptotically converge to such a behavior.

%
\begin{figure}
	\centering
	\includegraphics[width=0.9\linewidth]{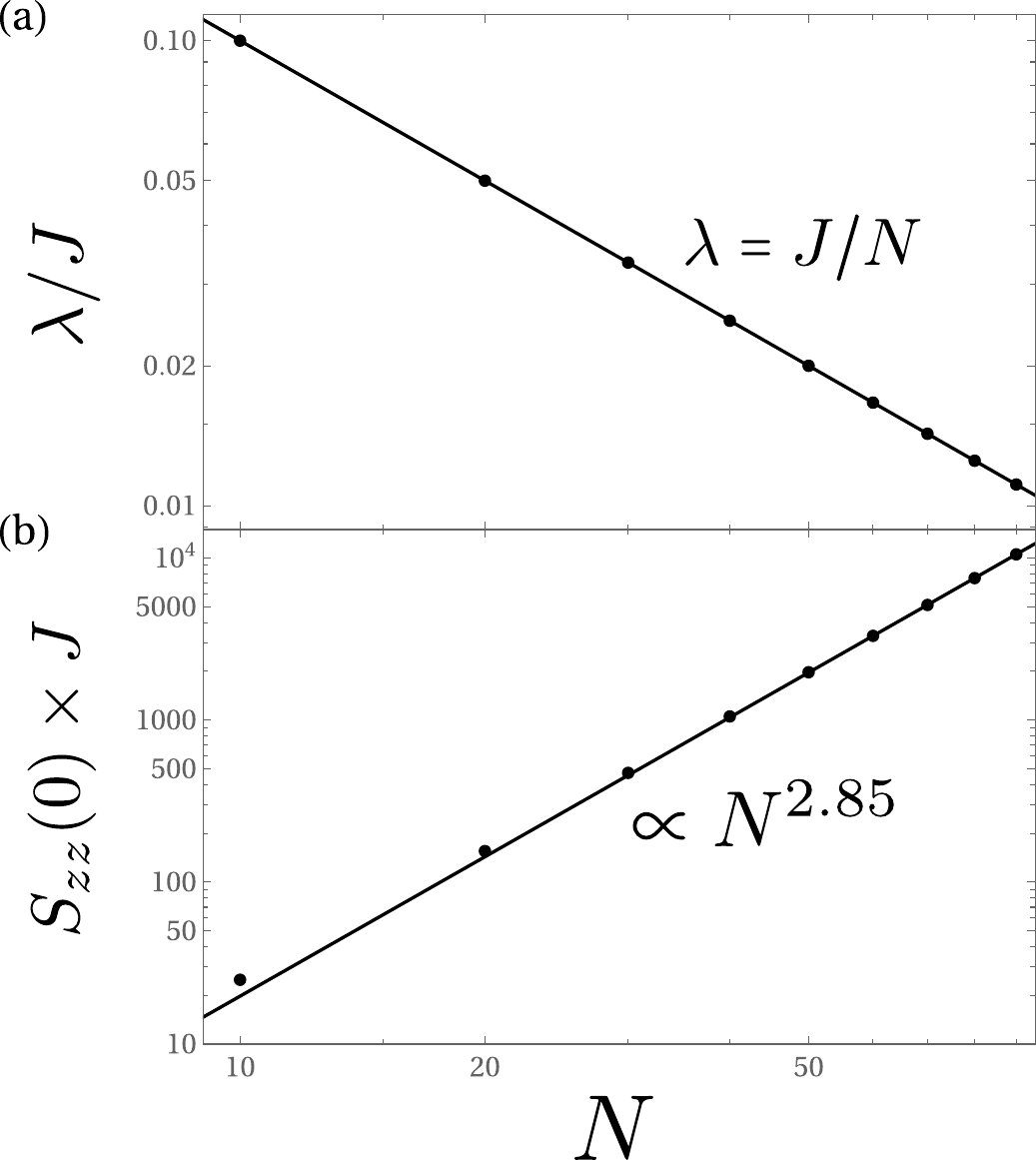}
	\caption{The finite-size scaling with $N$ of the Liouvillian gap (a) and fluctuations (b) for $\gamma=\gamma_c$, plotted in the log-log scale. The results are denoted by dots, black solid lines correspond to power-law fits for large $N$.}
	\label{fig:gapfluctlmg}
\end{figure}
In Fig.~\ref{fig:gapfluctlmg} we present the finite-size scaling of the Liouvillian gap and the zero-frequency noise $S_{zz}(0)$ at the phase transition point $\gamma=\gamma_c$. Both quantities exhibit a power-law scaling; intriguingly, the Liouvillian gap appears to be exactly given by the formula $\lambda=J/N$. This is yet another example of such behavior for systems exhibiting a discontinuous phase transition without phase coexistence, this time in a model with time-dependent attractors. As in the previous example, the scaling behavior can be explained by the diffusion along and between the periodic orbits.

%
\begin{figure}
	\centering
	\includegraphics[width=0.9\linewidth]{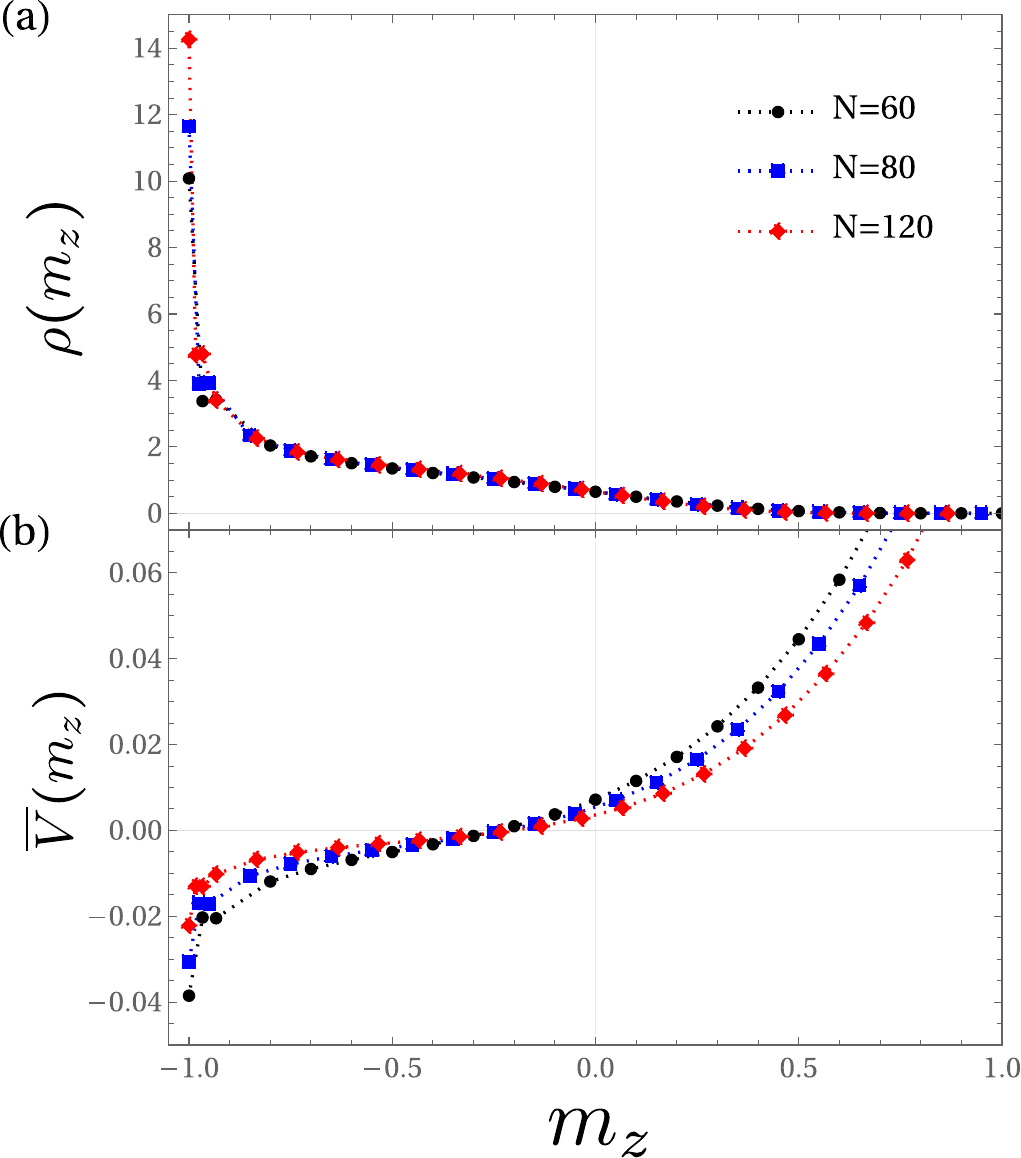}
	\caption{The probability density $\rho(m_z)$ (a) and the estimator of the rate function $\overline{V}(m_z)$ (b) for $\gamma=\gamma_c$.}
	\label{fig:densratelmg}
\end{figure}
Finally, let us consider the properties of the probability distribution of the system magnetization $m_z$ at the phase transition point. We recall that the probability distribution $P(q)$ for the molecular zipper with $a=0$ and the distribution $P(m_z)$ for the quantum Ising model were uniform at discontinuous phase transitions without phase coexistence. In the case analyzed now, this is no longer true. However, we show that the probability distribution becomes uniform in the large deviation sense, i.e., the rate function becomes constant, which corresponds to the presence of infinitely many attractors. According to large deviation theory, in the large $N$ limit, the probability density $\rho(m_z)=N P(m_z)$ of the system magnetization is described by the asymptotic expression~\cite{FalascoReview}
\begin{align}
\rho(m_z) \asymp A(m_z) e^{-N V(m_z)},
\end{align}
where, in addition to the rate function $V(m_z)$ defined in Sec.~\ref{sec:disc}, we introduce the scale-independent subexponential prefactor $A(m_z)$. We can further define the estimator of the rate function for a finite $N$,
\begin{align}
\overline{V}(m_z) \equiv -N^{-1} \ln \rho(m_z),
\end{align}
which converges to $V(m_z)$ for $N \rightarrow \infty$. In Fig.~\ref{fig:densratelmg} we plot $\rho(m_z)$ and $\overline{V}(m_z)$ at the phase transition point for different system sizes. As shown, with increasing system size, the probability density becomes approximately scale-independent (with certain deviations in the vicinity of $m_z=-1$). The estimator $\overline{V}(m_z)$, instead, gradually approaches zero in the whole domain of $m_z$. Thus, one may expect that in the large $N$ limit the rate function becomes constant ($V(m_z)=0$), and the nonuniformity of the probability density is only determined by the subexponential prefactor: $\rho(m_z) \asymp A(m_z)$. The latter is determined by the drift term, which determines the time spent in different segments of periodic orbits, as well as by the diffusive term, which determines the diffusion between orbits. An analogous behavior has been observed for the cooperative fluorescence model that undergoes a continuous phase transition to the phase with infinitely many periodic orbits~\cite{drummond1978volterra,walls1978non,carmichael1980analytical,barberena2023critical}.

\section{Steady state degeneracy: Liouvillian gap closing for finite systems} \label{sec:sdm}
So far we have focused on the cases where the Liouvillian has a unique steady state for finite system sizes, and thus the Liouvillian gap closes only asymptotically in the thermodynamic limit. Here, at the end of our discussion, we note another possibility in which the steady state becomes degenerate at the phase transition point, and thus the Liouvillian gap closes for any system size. A notable example is the squeezed decay model, which has been thoroughly studied in Refs.~\cite{agarwal1989nonequilibrium, agarwal1990cooperative,barberena2023critical}. It is described by the master equation
\begin{align}
\dot{\rho}=\frac{\gamma}{N} \mathcal{D}[\hat{S}_x-i \zeta \hat{S}_y](\rho),
\end{align}
where $\gamma$ is the dissipation rate, and $\zeta \in [-1,1]$ is the squeezing parameter. The model can be experimentally realized using an ensemble of $N$ atoms irradiated by the broadband squeezed light~\cite{agarwal1989nonequilibrium,agarwal1990cooperative} or subjected to two-photon Raman transitions~\cite{dalla2013dissipative}.

\begin{figure}
	\centering
	\includegraphics[width=0.9\linewidth]{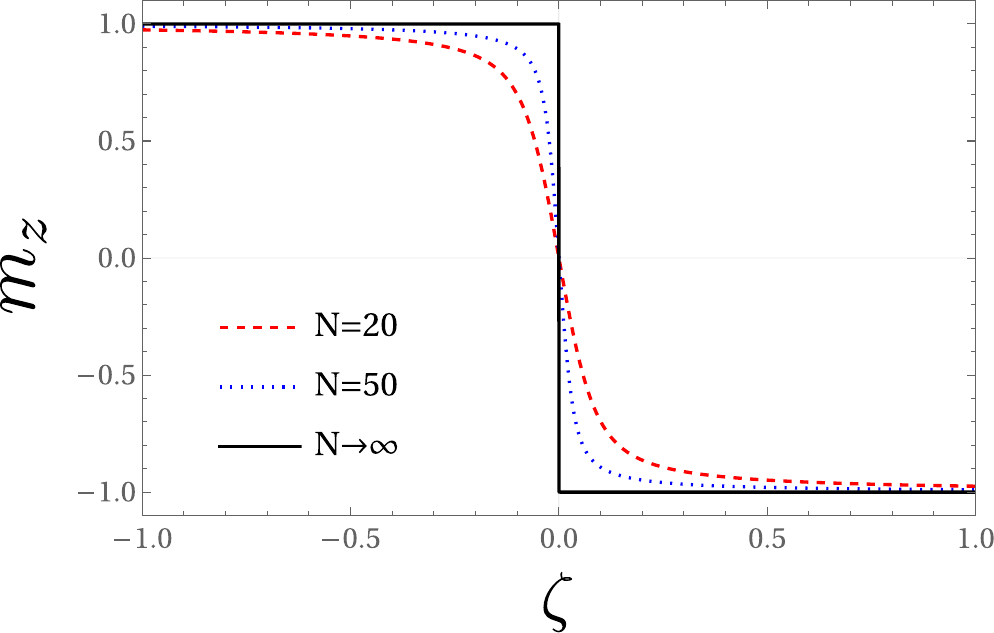}
	\caption{The normalized magnetization $m_z$ in the squeezed decay model as a function of the squeezing parameter $\zeta$ for different values of $N$.}
	\label{fig:squeezeddec}
\end{figure}
The steady state of this model has been determined exactly. In particular, for even $N$, the normalized magnetization $m_z \equiv 2\langle \hat{S}_z \rangle/N$ can be expressed by the exact formula~\cite{barberena2023critical}
\begin{align}
m_z=-\frac{I_1(\zeta N)}{I_0(\zeta N)},
\end{align}
where $I_k$ is the $k$th modified Bessel function of the first kind. For odd $N$, this formula is approximate, with accuracy increasing with $N$. The dependence of $m_z$ on $\zeta$ for different values of $N$ is plotted in Fig.~\ref{fig:squeezeddec}. As shown, in the thermodynamic limit $N \rightarrow \infty$ the system exhibits a discontinuous phase transition at $\zeta=0$, with the normalized magnetization jumping from $m_z=1$ to $m_z=-1$. We further note that this transition is not associated with phase coexistence. Indeed, the system can be described by the same mean field theory as in Sec.~\ref{sec:ising}, with $h=J=0$ and $\gamma \rightarrow \zeta \gamma$. Therefore, it has a unique attractor $m_z=1$ ($m_z=-1$) for $\zeta<0$ ($\zeta>0$),

At the same time, it is easy to see that for $\zeta=0$ the Liouvillian commutes with the $\hat{S}_x$ operator. Therefore, the system dynamics is not irreducible, since each of the $N+1$ eigenstates of $\hat{S}_x$ is a steady state of the system. Qualitatively, based on microscopic derivations of the model~\cite{agarwal1989nonequilibrium,agarwal1990cooperative,dalla2013dissipative}, this can be interpreted as a result of the suppression of quantum jumps between eigenstates of $\hat{S}_x$ due to negative quantum interference. As a result, the system does not relax to a unique stationary state but rather undergoes a pure dephasing to the statistical mixture of eigenstates of $\hat{S}_x$ that depends on initial conditions. Consequently, the dominant eigenvalue of the Liouvillian is $(N+1)$-fold degenerate: $\lambda_0=\lambda_1 =\ldots= \lambda_N=0$. Therefore, the Liouvillian gap exhibits no finite-size scaling but is instead equal to 0 at the phase transition point for any system size. Furthermore, since the stationary state is not unique, stationary fluctuations are not well defined, which is called the zero-frequency anomaly~\cite{hanggi1982stochastic}.

\section{Conclusions} \label{sec:concl}
We have shown that the finite-size scaling of the Liouvillian gap and dynamical fluctuations at discontinuous phase transitions depends on whether the transition is associated with the phase coexistence or not. In the former case, when the phase transition point is surrounded by a finite phase coexistence region, one observes an exponential scaling, related to stochastic switching between the system attractors. In the latter case, where the phase transition point separates the phase diagram regions with distinct attractors, the Liouvillian gap and fluctuations exhibit instead a power-law scaling, related to the diffusive nature of the system relaxation to the stationary state. Thus, the exponential scaling of the Liouvillian gap and fluctuations is associated with phase coexistence rather than with the discontinuous nature of the phase transition itself.

Furthermore, even in the presence of phase coexistence, the scaling of the Liouvillian gap and fluctuations exhibits actually a crossover from power-law to exponential scaling, rather than a purely exponential behavior. This is most apparent when the coexisting attractors are relatively strongly coupled (e.g., when the free energy density barrier between them is relatively small). In such a case, the onset of exponential scaling may occur only for very large system sizes. This may hinder the observability of exponential scaling in experiments and simulations, which are often confined to relatively small system sizes. Consequently, this may explain the observations of power-law scaling in certain simulations~\cite{vicentini2018critical} and experiments~\cite{sett2024emergent}.

Finally, we note that, in some cases, the Liouvillian gap can close at the discontinuous phase transition point even for finite system sizes. This occurs when the steady state becomes degenerate at this point. We illustrate this on the squeezed decay model discussed in Sec.~\ref{sec:sdm}.

\begin{acknowledgments}
K.P.\ acknowledges the financial support of the National Science Centre, Poland, under the project No.\ 2023/51/D/ST3/01203.
\end{acknowledgments}

\appendix

\section{Derivation of Eqs.~\eqref{arrh} and~\eqref{arrhamp}} \label{app:derexpgap}
Here we derive the Arrhenius-like formula for the Liouvillian gap [Eq.~\eqref{arrh}], with the pre-exponential factor given by Eq.~\eqref{arrhamp}. To do so, we use the approach presented in Ref.~\cite{hinch2005exponentially}. Let us first recall the rate function $V(q)=\beta \mathcal{F}(q)$ defined below Eq.~\eqref{eq:largedevzipp} (we note that in Ref.~\cite{hinch2005exponentially} the rate function was defined with an opposite sign). At the critical temperature $T_c$ given by Eq.~\eqref{eq:dynvecdec} it takes the form
\begin{align} \label{largedevcrit}
		V(q)= \frac{a(q-q^2)}{1+a} \ln g - \ln g.
\end{align}
We denote the positions of the minima of $V(q)$ as $q_-$ and $q_+$, while the position of the maximum as $q_\text{max}$. For the model considered $q_-=0$, $q_+=1$, and $q_\text{max}=1/2$. The asymptotic value of the Liouvillian gap can then be found by equating Eq.~(4.14) and~(4.15) from Ref.~\cite{hinch2005exponentially}, which yields
\begin{align} \label{hincheq}
\lambda \approx e^{-NV(q_\text{max})} \sqrt{\frac{-\gamma_0}{2\pi N}} \left(\frac{1}{\mathcal{T}_-}+\frac{1}{\mathcal{T}_+} \right),
\end{align}
where
\begin{align} \label{eq:tpm}
\mathcal{T}_\pm &\approx \pm N \int_{q_\text{max}}^{q_\pm} dq \frac{e^{-N V(q)}}{\sqrt{w_+(q) w_-(q)}}, \\
\gamma_0&= \frac{w'_-(q_\text{max})-w_+'(q_\text{max})}{w_+(q_\text{max})}.
\end{align}
The rates $w_\pm (q)$ given by Eq.~\eqref{eq:zipprates} are evaluated at $T=T_c$. The term $\gamma_0$ takes then the form
\begin{align} \label{eq:gamma0}
	\gamma_0=-\frac{2 a \ln g}{1+a}.
\end{align}
To calculate the terms $\mathcal{T}_\pm$, we use the Laplace method. We note that for large $N$ the integral of the exponent $\exp[-NV(q)]$ is dominated by the region where $V(q)$ is smallest, that is, close to $q_\pm$. In this region, the function $V(q)$ can be expanded as:
\begin{align}
	V(q)=V(q_\pm)+V'(q_\pm)(q-q_\pm)+O[(q-q_\pm)^2],
\end{align}
with $V'(q_\pm)=\mp a \ln g/(1+a)$. Here we apply linear rather than quadratic expansion, as in the standard Laplace method, because the minima of $V(q)$ are located at the boundaries of its domain, where the derivatives $V'(q)$ do not disappear~\cite{hinch2005exponentially}. Inserting this into Eq.~\eqref{eq:tpm} one gets (in the large $N$ limit)
\begin{align} \label{eq:tpm-comp}
	\mathcal{T}_+=\mathcal{T}_- &\approx \frac{1+a}{a} \frac{e^{-N V(q_-)}}{\Gamma \sqrt{g} \ln g}.
\end{align}
Finally, inserting Eqs.~\eqref{eq:gamma0} and~\eqref{eq:tpm-comp} into Eq.~\eqref{hincheq}, we obtain the expression for the Liouvillian gap:
\begin{align}
\lambda \approx \frac{2 \Gamma \sqrt{g}}{\sqrt{\pi N}} \left( \frac{a \ln g}{1+a} \right)^{3/2} e^{-N[V(q_\text{max})-V(q_-)]}.
\end{align}
This corresponds to Eq.~\eqref{arrh}, with the pre-exponential factor $A$ given by Eq.~\eqref{arrhamp}.

\bibliography{bibliography}	
	
\end{document}